\documentclass[%
 aip,
 amsmath,amssymb,
 reprint,%
]{revtex4-1}

\usepackage{graphicx}
\usepackage{dcolumn}
\usepackage{bm}

\usepackage[utf8]{inputenc}
\usepackage[T1]{fontenc}
\usepackage{mathptmx}
\usepackage{etoolbox}
\usepackage{float}

\makeatletter
\def\@email#1#2{%
 \endgroup
 \patchcmd{\titleblock@produce}
  {\frontmatter@RRAPformat}
  {\frontmatter@RRAPformat{\produce@RRAP{*#1\href{mailto:#2}{#2}}}\frontmatter@RRAPformat}
  {}{}
}%
\makeatother

\usepackage[usenames]{color}

\usepackage[normalem]{ulem} 

\begin{document}

\preprint{AIP/123-QED}

\title[Phase boundaries and the Widom line from the Ruppeiner geometry of fluids]{Phase boundaries and the Widom line from the Ruppeiner geometry of fluids}

\author{Karlo de Leon}
\affiliation{ 
Department of Physics, New York University, New York, NY 10003, USA
}
\email{knd8618@nyu.edu}
\author{Ian Vega}%
\affiliation{ 
National Institute of Physics, University of the Philippines, Diliman, Quezon City 1101, Philippines
}%


\date{\today}

\begin{abstract}
In the study of fluid phases, the Ruppeiner geometry provides novel ways for constructing the phase boundaries and the Widom line of a thermodynamic state space. In this paper, we revisit these geometry-based constructions with the aim of understanding their limitations and generality. Our analysis is based in part on a new equation-of-state expansion for fluids near a critical point that assumes analyticity with respect to the number density. This parametrization is used to prove the equivalence of the geometric method and the standard Maxwell construction of phase boundaries near the critical point for a broad class of fluids. In contrast, the geometric and standard thermodynamic constructions of the Widom line are found to be generally inequivalent. But for the van der Waals fluid, we introduce a small reformulation of the Ruppeiner metric, which we call the Ruppeiner-$N$ metric, that makes the agreement between constructions exact. This improves on the results of May and Mausbach who find agreement only up to the slope of the Widom line when using the usual Ruppeiner metric. We also find that the proposed Ruppeiner-$N$ metric improves the classification scheme of Diósi \textit{et al.} for partitioning the van der Waals state space into its different phases using Ruppeiner geodesics. Whereas the original Diósi boundaries do not correspond to any established thermodynamic lines above (or even below) the critical point, the boundaries of Ruppeiner-$N$ geodesics are able to detect the presence of the Widom line. These results suggest that the Ruppeiner-$N$ metric may be the more appropriate metric to use when studying phase diagrams with thermodynamic geometry.
\end{abstract}

\maketitle


\section{\label{sec:intro}Introduction}

The infusion of geometric ideas in thermodynamics traces a long illustrious history.\cite{gibbs1948collected,caratheodory1909untersuchungen,hermann1973geometry,mrugala1978geometrical,mrugala1985submanifolds,weinhold1975metric,ruppeiner1979thermodynamics} In one sense, this is quite natural. Already, one can see intimations of geometry in determining the stability of thermodynamic states. Stability demands the convexity or concavity of fundamental-relation surfaces, depending on what thermodynamic potential is used. In the entropy representation, the entropy $S$ is given as a function of the extensive variables $x^i$, where $x^0=E$ (energy), $x^1=V$ (volume), etc. An equilibrium state on the hypersurface $S=S(x^i)$ is stable when the surface at that point is concave with respect to any direction on the hyperplane $(x^i)$.\cite{callen1985thermodynamics} This concavity of the entropy is encoded in its stability matrix, or its Hessian, $D^2 S$. The significance of a (proto-)geometric property, the concavity, in the thermodynamics of systems has motivated explorations of other geometric properties of thermodynamics that have broad physical significance. It remains a fascinating question just how much of thermodynamics can be viewed from the lens of geometry, and what applications are enabled by such a geometric viewpoint. 

One complication towards this goal is that there is no unique or natural metric for thermodynamic geometry. Gibbs\cite{gibbs1948equilibrium} himself was already skeptical of the uniqueness of such a metric, as was Tisza\cite{tisza1966generalized}. Instead, different metrics for thermodynamic spaces have been proposed in the literature, depending on the application and on what aspects of thermodynamics one wishes to highlight. The earliest proposed metric appears to be that of Weinhold\cite{weinhold1975metric}, who defined it to be the Hessian of the internal energy $D^2 E$. Weinhold showed that the laws of thermodynamics can be restated as mathematical statements requiring the Weinhold metric to be Riemannian. Closely related to this, though coming from an entirely different motivation, is the Ruppeiner metric\cite{ruppeiner1979thermodynamics}. This metric is instead the negative Hessian of the entropy $-D^2 S$ whose interpretation is rooted in thermodynamic fluctuation theory. These two metrics were later shown\cite{salamon1984relation} to be conformal to each other through a factor of temperature: $D^2 E = T(-D^2 S)$. In this study, we focus on the Ruppeiner metric.

Since the introduction of the Ruppeiner metric in 1979\cite{ruppeiner1979thermodynamics}, a substantial body of work\cite{ruppeiner1983new,ruppeiner1995riemannian,ruppeiner2010thermodynamic,ruppeiner2012thermodynamiccrit} has been directed towards firming up its theoretical groundwork as well as developing its potential uses. The Ruppeiner metric has found many important applications in thermal physics, such as in fluctuation theory\cite{ruppeiner2010thermodynamic}, finite-time thermodynamics\cite{feldmann1985thermodynamic,nulton1985quasistatic,salamon1985length}, phase transitions \cite{ruppeiner2012thermodynamiccrit,ruppeiner2012thermodynamic,may2012riemannian,may2013thermodynamic,ruppeiner2015thermodynamic,jaramillo2019r,castorina2018thermodynamic,wei2019ruppeiner,chaturvedi2017thermodynamic}, and even black hole thermodynamics\cite{ruppeiner2018thermodynamic,wei2019ruppeiner,chaturvedi2017thermodynamic}, among others. The interest of the present work is on phase transitions.

Given a metric, the immediate objects of interest are its geodesics and curvature. Much of the work\cite{feldmann1985thermodynamic,nulton1985quasistatic,salamon1985length,ruppeiner2012thermodynamiccrit,ruppeiner2012thermodynamic,may2012riemannian,may2013thermodynamic,ruppeiner2015thermodynamic,jaramillo2019r,castorina2018thermodynamic,wei2019ruppeiner,chaturvedi2017thermodynamic,ruppeiner2018thermodynamic,wei2019ruppeiner,chaturvedi2017thermodynamic} on the Ruppeiner metric is anchored on these two geometric objects, though most papers tend to deal with either one or the other. In the present work, we shall explore both. 


We focus first on geodesics, and begin by mentioning three physical interpretations that have been proposed for them. The geodesic from state $A$ to state $B$ in the Ruppeiner geometry can be thought of as: (1) the most probable path that fluctuations can drive the system from $A$ to $B$, which derives from fluctuation theory\cite{ruppeiner1979thermodynamics,ruppeiner2010thermodynamic}, (2) the path of minimum number of distinguishable fluctuations between $A$ and $B$,\cite{diosi1984metricization}, or 
(3) the path of least dissipation in a finite-time process\cite{feldmann1985thermodynamic,nulton1985quasistatic,salamon1985length}. One interesting study by Diósi et al.\cite{diosi1989mapping} on the van der Waals fluid explored the use of geodesics in determining whether two thermodynamic states belong to the same phase, and thereby sought to provide a definition for phases based on geodesics. In the standard analysis, the phases of two distinct states below the critical point are operationally defined in terms of isotherms and isobars. In the language of differential geometry, however, this would be tantamount to singling out special coordinate systems, which would be anathema to a geometric treatment of thermodynamics. Of course, the temperature and pressure are quantities already imbued with physical meaning, not like the $x$ and $y$ axes of the Cartesian plane which are merely coordinates. But in thermodynamic geometry, variables such as the temperature and pressure are to be viewed as just two of the many coordinates that can chart the thermodynamic state space. The main advantage of using invariant geometric objects like geodesics in classifying the phases of thermodynamic states is that the scheme  will remain valid in any coordinate representation of the state space. With this in mind, Diósi et al. introduced a partitioning of the van der Waals space that separates ``gas'' and ``liquid'' regions even above the critical point. However, as we will discuss later, their classification scheme suffers from a key weakness: their region boundaries tend not to correspond to any physically meaningful thermodynamic line. 

In this paper, we discover that by slightly reformulating the Ruppeiner metric, we can turn the Diósi et al. geodesic-based partitioning scheme into one in which the geodesics are able to detect the presence of the Widom line, a thermodynamic line considered by many\cite{mcmillan2010going,simeoni2010widom,xu2005relation} to be the continuation of the coexistence curve on the supercritical region of a fluid. The small change to the Ruppeiner metric is quite simple. In the extant literature, the Ruppeiner metric is computed with the volume set constant. According to Ruppeiner's exposition\cite{ruppeiner1995riemannian}, a constant-volume analysis allows for ready interpretation of quantities, such as the Ricci curvature being correlated with the correlation length. Here, we allow the volume to change and keep the number of particles constant instead. We call the original formulation the \textit{Ruppeiner-V metric}, and our proposed formulation the \textit{Ruppeiner-N metric}. We will see that the Ruppeiner-$N$ metric brings interesting results for some systems and can provide better results for curves like the Widom line than the conventional Ruppeiner-$V$ metric.

The term ``Widom line'' is used rather loosely by different authors. Originally, it was defined to be the locus of points around a critical point that maximize the correlation length.\cite{mcmillan2010going,xu2005relation} Now, whether this maximum is taken along isobars\cite{ruppeiner2012thermodynamic,xu2005relation,banuti2015crossing,ruppeiner2015thermodynamic} or isotherms\cite{mcmillan2010going,simeoni2010widom,brazhkin2011widom} tends to be ambiguous in the literature. Since correlation lengths are challenging to measure or compute, the Widom line is often obtained indirectly through the maxima of thermodynamic response functions instead, such as the isobaric heat capacity, isothermal compressibility, or the thermal expansion coeﬀicient.\cite{mcmillan2010going,may2012riemannian} The reason behind this is that all these response functions scale as powers of the correlation length near the critical point. So, the curves of these response function maxima should all asymptote to the Widom line at the critical point. While this method is disputed by some authors\cite{banuti2015crossing,may2012riemannian}, it has become the established technique in obtaining the Widom line, with the isobaric heat capacity most often used as the response function to approximate the correlation length\cite{ruppeiner2012thermodynamic,banuti2015crossing,lamorgese2018widom}. Due to this, some authors\cite{lamorgese2018widom} refer to the Widom line directly as the locus of points that maximize the isobaric heat capacity. To differentiate these two definitions, we refer to the correlation-length-based curve as the \textit{statistical} Widom line (since the correlation length is a statistical quantity), and the heat-capacity-based curve as the \textit{thermodynamic} Widom line (in the same spirit, since the heat capacity is a thermodynamic quantity). Furthermore, we may add the qualifiers ``isobar'' and ``isotherm'' to clarify along what paths the maxima are taken. The Widom line gained the attention of the scientific community when various systems were observed to undergo abrupt changes in its properties when crossing the purported line.\cite{xu2005relation,nishikawa1995correlation,banuti2015crossing,gorelli2006liquidlike} In some supercritical fluids, the system shows liquid-like properties on one side of the line and gas-like properties on the other.\cite{simeoni2010widom} This has led some authors\cite{mcmillan2010going,simeoni2010widom,xu2005relation} to consider the Widom line as the continuation of the coexistence curve into the supercritical region. 

Beyond geodesics, the curvature of the Ruppeiner metric has also been proposed to hold thermodynamic significance. There have many investigations \cite{ruppeiner2010thermodynamic} of the Ricci scalar curvature of the Ruppeiner metric for different thermodynamic systems. The sign of the Ricci scalar is seen to reflect the nature of the dominant intermolecular forces in a system. For example, the curvature of the simple ideal gas vanishes everywhere, whereas the Fermi and boson gas have positive and negative curvature values, respectively.\cite{ruppeiner2010thermodynamic} A list of the Ricci scalar signs of different systems can be found in Ref.~\onlinecite{ruppeiner2010thermodynamic}. The physical interpretation of the magnitude of the Ricci scalar is a popular topic for much of the research concerning the Ruppeiner metric. Ruppeiner\cite{ruppeiner1979thermodynamics,ruppeiner2010thermodynamic,ruppeiner2012thermodynamic,ruppeiner2012thermodynamiccrit} himself hypothesized that the magnitude of the Ricci scalar is proportional to the correlation volume near the critical point \begin{equation}
    R \propto \xi^3.\label{eq:Rpropto}
\end{equation}
This relation holds for many systems\cite{ruppeiner1979thermodynamics,ruppeiner1981application,ruppeiner1990thermodynamic,brody1995geometrical,dolan1998geometry,dolan2002information,janke2002information,janke2003information,brody2003information,johnston2003information}, though it has not yet been rigorously proven to be generally true for all systems. Notably, the Ricci scalar and the correlation length both diverge at the critical point, and both are found to have the same critical exponent\cite{ruppeiner2012thermodynamiccrit}. 

Ruppeiner's hypothesis opens up a number of possible applications for the scalar curvature. In this paper, we pay attention to two of them. By Widom's argument  that the correlation lengths of coexistent states are equal,\cite{widom1974critical} Ruppeiner's hypothesis implies that the Ricci curvature of coexistent states should also be equal. This enables a geometric construction of phase boundaries using the Ricci curvature. That is, in contrast to the standard analysis in which we equate the Gibbs free energy of two coexistent states, in a geometry-based construction we equate their Ricci curvatures instead\cite{ruppeiner2012thermodynamic}. We call the boundary generated using standard methods the \textit{Maxwell} phase boundary (from Maxwell's equal area law), and the Ricci-based boundary the \textit{Ricci} phase boundary. This alternative method of obtaining the phase boundary is sometimes referred to as the $R$\textit{-crossing method}. A number of authors\cite{ruppeiner2012thermodynamic,may2012riemannian,jaramillo2019r,castorina2018thermodynamic,may2013thermodynamic,wei2019ruppeiner,chaturvedi2017thermodynamic} have applied the method to different systems. In general, excellent agreement is found between the two phase boundaries \emph{near the critical point}. One of the key motivations of the present work is to understand if this is generally true and, if so, then why. 

In order to do this, we construct a general parametrized expansion of fluid equations-of-state (EoS) near the critical point, and then calculate the associated thermodynamic metrics and their curvature. We then use this to assess the generality of the $R$-crossing method. Based on this, we provide an explicit proof that the Maxwell and Ricci phase boundaries must agree near the critical point for a broad class of fluids.

Other than providing an alternative means for constructing phase boundaries, the Ricci scalar is also expected to generate a (statistical) Widom line, as per Ruppeiner's hypothesis in Eq. \eqref{eq:Rpropto}. We call this the \textit{Ricci-Widom line}.
May and Mausbach\cite{may2012riemannian} calculated the Ricci-Widom line of the van der Waals fluid and compared it to its thermodynamic Widom line; the curves were seen to have equal slopes at the critical point. In this paper, we show that if we use the Ruppeiner-$N$ metric instead, we will get even much better agreement: the van der Waals thermodynamic Widom line and Ricci-Widom line are exactly the same, even when they are far from the critical point.

Summarizing what lies ahead, this paper proposes the use of the Ruppeiner-$N$ metric for the characterization of phases in thermodynamic systems, as opposed to the Ruppeiner-$V$ metric that has been universally used thus far. The geodesics of the Ruppeiner-$N$ metric give rise to a more physically meaningful phase-partitioning scheme than the one proposed by Di\'{o}si et al, one that is sensitive to the Widom line. Moreover, the Ruppeiner-$N$ curvature provides a number of key results. We are able to prove that for a general class of fluids the phase boundaries constructed based on this curvature (i.e., the $R$-crossing method) are identical to those based on the Maxwell construction. However, it will be shown later that there are infinitely many construction schemes that can also do this, and so this is not unique to the $R$-crossing method. Nevertheless, this tells us that the $R$-crossing method is consistent with the standard results of thermodynamics. We also show that this curvature allows for a construction of a Ricci-Widom line that exactly matches the thermodynamic Widom line for the van der Waals fluid. These results suggest that there may be a greater role to be played by the Ruppeiner-$N$ metric in applications of thermodynamic geometry.

The paper is organized as follows. In Section \ref{sec:Rupp}, we review the fundamentals of the Ruppeiner metric. We then proceed to Section \ref{sec:geod} to discuss the application of the Ruppeiner-$N$ metric in characterizing the phases of the van der Waals fluid. We also give a review to the study of Diósi et al.\cite{diosi1989mapping} regarding this partitioning scheme. In Section \ref{sec:expand}, we present the parametrized fluid EoS expansion we designed specifically for Ruppeiner metric applications. These results are used in Section \ref{sec:ricci} where we investigate the relation of the Ruppeiner curvature to a system's phase boundary and Widom line. Finally, we conclude the paper in Section \ref{sec:conclusion}.

\section{\label{sec:Rupp}Ruppeiner metric}

\subsection{Thermodynamic fluctuation theory}

The Ruppeiner metric naturally appears in thermodynamic fluctuation theory. Consider a system in contact with its environment and whose equilibrium macrostate is at $A_0$. With the assumption of equal microstate probabilities, the probability of the system to be at a macrostate $A$ must be proportional to the number of microstates of $A$:
\begin{equation}
    P_A \propto \Omega_A.
    \label{eq:prob}
\end{equation}
By inverting Boltzmann's equation $S = k \ln{\Omega}$, we can write this probability in terms of thermodynamic quantities:
\begin{equation}
    P_A \propto \exp{(S^U_A/k)},
    \label{eq:prob_S}
\end{equation}
where $S^U_A$ is the entropy of the universe when the system is at state $A$, and $k$ is the Boltzmann constant. Equation \eqref{eq:prob_S} is Einstein's famous relation in fluctuation theory.\cite{landau1980statistical} To incorporate the phenomenon of fluctuation in our analysis, we expand the entropy about the equilibrium state $A_0$:
\begin{equation}
    S^U = \sum_{n=0}^{\infty}\frac{1}{n!} \Big[(\textrm{d}\mathbf{X}\cdot\mathbf{\nabla})^n_{A_0} \,S + (\textrm{d}\mathbf{X'}\cdot\mathbf{\nabla})^n_{A_0} \,S' \Big],
    \label{eq:S_expand}
\end{equation}
where the $X^\mu$ components of $\mathbf{X}$ are the natural extensive variables of the entropy. Unprimed variables denote system quantities while primed variables denote environment quantities. Because $A_0$ is the equilibrium state, the $n=1$ terms of Eq. \eqref{eq:S_expand} are all zero. Up to the non-vanishing leading order, the entropy of the universe is
\begin{equation}
    S^U = S^U_{A_0} + \Big(\frac{1}{2} (\mathbf{\textrm{d}X}\cdot\mathbf{\nabla})^2_{A_0} \,S + \frac{1}{2} (\mathbf{\textrm{d}X'}\cdot\mathbf{\nabla'})^2_{A_0} \,S'\Big).
    \label{eq:S_expand_last}
\end{equation}
The Hessians of $S$ and $S'$ already appear in the second and third term of Eq. \eqref{eq:S_expand_last}, respectively. To set the scale of the system, we keep one extensive variable constant. Let this scale variable be $X^{\textrm{sc}}$. This also fixes the corresponding environment extensive variable $X'^{\textrm{sc}}$. We then take the densities of the extensive variables with respect to scale variables, i.e., $x^\mu:=X^\mu/ X^{\textrm{sc}}$ and $x'^\mu:=X'^\mu/ X'^{\textrm{sc}}$. Because extensive quantities are conserved for a closed system such as the universe, ${\textrm{d}x'}^\mu = - \epsilon \,{\textrm{d}x}^\mu$, where $\epsilon = X^{\textrm{sc}}/X'^{\textrm{sc}}$. The last term in Eq. \eqref{eq:S_expand_last} is of the order $\epsilon^2$. Assuming that the environment is very large compared to the system, in the sense that $X^{\textrm{sc}} \ll X'^{\textrm{sc}}$ ($\epsilon \ll 1$), we ignore non-linear terms in $\epsilon$. Finally, dropping the last term of Eq. \eqref{eq:S_expand_last} and plugging this expression of the entropy to Eq. \eqref{eq:prob_S} we get
\begin{equation}
    P_A = C\exp{ \Big[ -\frac{1}{2k} \Big(-\frac{\partial^2 S}{\partial X^\alpha \partial X^\beta}\Big)_{A_0} \delta X^\alpha_A \delta X^\beta_A \Big]},
    \label{eq:prob_S_beforefinal}
\end{equation}
where $C$ is the normalization constant, and $\delta X^\alpha_A := X^\alpha_A-X^\alpha_{A_0}$. We are using Einstein's summation convention for repeated indices. Ruppeiner\cite{ruppeiner1979thermodynamics,ruppeiner2010thermodynamic} notes that the Hessian of the system entropy qualifies as a Riemannian metric and defines this to be the Ruppeiner metric
\begin{equation}
    g_{\alpha\beta} = -\frac{\partial^2 S}{\partial X^\alpha \partial X^\beta}.
    \label{eq:Rupp}
\end{equation}
In the literature, $X^{\textrm{sc}}$ is usually chosen to be the volume $V$ of the system; we call this the Ruppeiner-$V$ metric. But here, we we will work with the number of system particles $N$ to be the scale variable which we call the Ruppeiner-$N$ metric. The choice of choosing $N$ to be constant is to parallel the methods in phase transitions analysis, where $N$ is also set to be constant.

Writing Eq. \eqref{eq:prob_S_beforefinal} as
\begin{align}
    P_A &= C\exp{ \Big( -\frac{1}{2k} g_{A_0} (\delta \mathbf{X}_A ,\delta\mathbf{X}_A ) \Big)}
    \label{eq:prob_S_final1}\\
    &=C\exp{ \Big( -\frac{1}{2k} |\delta \mathbf{X}_A|^2_{A_0}}\Big),
    \label{eq:prob_S_final2}
\end{align}
we see that the probability of the system to fluctuate from its equilibrium state $A_0$ to a nearby state $A$ is measured by the distance between $A$ and $A_0$ under the Ruppeiner metric; the smaller the distance, the larger the probability. Let us now consider the case when $A$ is distant from the equilibrium state $A_0$. We let the system go through a series of stepwise fluctuations that will bring the system from $A_0$ to $A$. The physical picture of this is that we have a series of baths with preset temperatures, pressures, etc. We immerse the system to one bath after another until it reaches the state $A$. Let the intermediate states be indexed by $t$ from $t=0$ to $t=\tau$. The total probability of going from $A_0$ to $A$ is then $P_{A_0\rightarrow A} = P_{A_0\rightarrow A_1}P_{A_1\rightarrow A_2}\ldots P_{A_{\tau-1}\rightarrow A_\tau}$ or
\begin{equation}
    P_{A_0\rightarrow A} = C\exp{ \Big( -\frac{1}{2k} \int_0^\tau g \big(\delta \mathbf{X}(t) ,\delta\mathbf{X}(t) \big)\, \textrm{d}t \Big)}.
    \label{eq:step}
\end{equation}
Invoking Titu's lemma\cite{sedrakyan1997applications}
\begin{equation}
    \frac{1}{N}\Big(\sum_{k=1}^N\,a_k\Big)^2\leq \sum_{k=1}^N\,a_k^2 ,\label{eq:titu}
\end{equation}
we find the upper bound of $P_{A_0\rightarrow A}$ to be
\begin{equation}
    P_{A_0\rightarrow A} \leq C\exp{ \Big[ -\frac{1}{2k} \Big( \int_0^\tau \sqrt{g \big(\delta \mathbf{X}(t) ,\delta\mathbf{X}(t) \big)}\, \textrm{d}t \Big)^2 \Big]}.
    \label{eq:upbound}
\end{equation}
The integral in Eq. \eqref{eq:upbound} is the distance of the path from $A_0$ to $A$. We see that the greater the distance, the smaller the upper bound of the probability of the system transitioning from state $A_0$ to $A$ becomes. In other words, states that are ``far'' in the thermodynamic space have low transition probabilities. Looking back at Eq. \eqref{eq:step}, notice that the probability is dependent on the parametrization of states. In finite-time thermodynamics\cite{salamon1983thermodynamic}, the parameter $t$ is interpreted as time. If the parametrization is such that $g \big(\delta \mathbf{X}(t) ,\delta\mathbf{X}(t) \big)$ is constant, we get the equality sign in Eq. \eqref{eq:upbound}. In summary, the Ruppeiner distance gives a measure of the probability of fluctuation from one state to another.

\subsection{Coordinates}

We are not constrained to express the Ruppeiner metric using the natural extensive variables of the entropy. Let $Y_\mu:=\partial S/\partial X^\mu$ be the conjugate variable of $X^\mu$. From Eq. \eqref{eq:Rupp}, we can write the Ruppeiner metric as
\begin{align}
    \label{eq:g_Rupp_dec1}
    g &= -\frac{\partial^2 S}{\partial X^\alpha \partial X^\beta} \textrm{d}X^\alpha \textrm{d}X^\beta\\
    &= -\textrm{d}X^\mu \textrm{d}Y_{\mu}.\label{eq:g_Rupp_dec2}
\end{align}
With the fundamental relation
\begin{equation}
    \textrm{d}E = T\textrm{d}S + P_i \textrm{d}X^i
    \label{eq:fundamental},
\end{equation}
where $P_i$ is the conjugate of $X^i$ in the energy representation, one can recast Eq. \eqref{eq:g_Rupp_dec2} as
\begin{equation}
    g = -\frac{1}{T}\textrm{d}T\,\textrm{d}S - \frac{1}{T} dX^i \textrm{d}P_i.
    \label{eq:useful}
\end{equation}
The Ruppeiner metric is diagonal in the coordinate system $(X^k,T)$ where $X^k$ is the lone extensive variable (aside from the energy and entropy) allowed to vary. In the $(N,T)$ frame where $V$ is constant, we have
\begin{equation}
    \label{eq:grupp_TN}
    g = \frac{1}{T}\Big( -\partial_T^2 F \, \textrm{d}T^2 + \partial_N^2 F \, \textrm{d}N^2\Big),
\end{equation}
where $F$ is the Helmholtz free energy. In the $(V,T)$ frame where $N$ is constant, we have
\begin{equation}
    \label{eq:grupp_TV}
    g = \frac{1}{T}\Big( -\partial_T^2 F \, \textrm{d}T^2 + \partial_V^2 F \, \textrm{d}V^2\Big).
\end{equation}
These two metrics in Eqs. \eqref{eq:grupp_TN} and \eqref{eq:grupp_TV} are different because we are looking at two different regions of the same thermodynamic manifold: one at the hyperplane of constant $V$, and one at the hyperplane of constant $N$. Equation \eqref{eq:grupp_TV} is the Ruppeiner-$V$ metric and Eq. \eqref{eq:grupp_TN} is the Ruppeiner-$N$ metric.

\subsection{Example: van der Waals fluid}

The simplest fluid that exhibits phase transitions is the van der Waals fluid\footnote{The simpler example of an ideal gas does not have any phase transitions and corresponds to a flat thermodynamic geometry. This has been studied in Ref. \onlinecite{nulton1985geometry}.}. Let us consider its Ruppeiner metrics. The ideal van der Waals fluid is characterized by the following pressure and energy EoS:\footnote{Ref. \onlinecite{callen1985thermodynamics}, p. 74.}
\begin{equation}\label{eq:P_EoS_vdw}
    P= \frac{NkT}{V-Nb} - a \frac{N^2}{V^2},
\end{equation}
\begin{equation}\label{eq:E_EoS_vdw}
    E= \frac{3}{2}NkT - a \frac{N^2}{V}.
\end{equation}
The corresponding Helmholtz free energy is
\begin{equation}\label{eq:free_vdw}
    F = -NkT \ln{\Big( \frac{V}{N} -b \Big)}-\frac{aN^2}{V}+h(T),
\end{equation}
where
\begin{equation}\label{eq:h}
    h(T) = -\frac{3}{2}NkT \ln{\Big( \frac{3}{2}kT\Big)} +Ts_0
\end{equation}
for some constant $s_0$. In our analysis, we always work in the frame $(x,T)$, where $x$ is the number density $x:=N/V$. It is also convenient to normalize quantities with respect to their critical values:
\begin{align}
    P_C &= \frac{1}{27}\frac{a}{b^2}\label{eq:crit_1}\\
    T_C &= \frac{8}{27}\frac{a}{bk}\label{eq:crit_2}\\
    x_C &= \frac{1}{3}\frac{1}{b}.\label{eq:crit_3}
\end{align}
From here, we will always use normalized quantities where all critical values are set to unity. For example, whenever the temperature ``$T$'' appears, we agree that it is a shorthand for $T/T_C$.

Evaluating Eqs. \eqref{eq:grupp_TN} and \eqref{eq:grupp_TV}, the Ruppeiner-$V$ and Ruppeiner-$N$ metric of the van der Waals fluid are
\begin{equation}\label{eq:g_vdw}
    g_{V} = \bigg(\frac{4x}{T^2}\bigg)\text{d}T^2 + \bigg(\frac{24}{x(x-3)^2} - \frac{6}{T}\bigg)\text{d}x^2,
\end{equation}
\begin{equation}\label{eq:g_vdw_N}
    g_{N} = \bigg(\frac{4}{T^2}\bigg)\text{d}T^2 + \frac{1}{x}\bigg(\frac{24}{x(x-3)^2} - \frac{6}{T}\bigg)\text{d}x^2,
\end{equation}
respectively. Note that the two metrics are conformal to each other: $g_N = \frac{1}{x}g_V$. We shall study these geometries extensively in what follows. 

\section{Thermodynamic geodesics\label{sec:geod}}

An immediate application of the Ruppeiner-$N$ metric is that its geodesics provide an improved coordinate-invariant partitioning scheme for defining phases in thermodynamic state spaces. But first we discuss some preliminaries.

The length $s$ of a curve $\gamma$ on a space with metric $g_{ab}$ is given by\cite{klingenberg2013course}
\begin{equation}
    s = \int_\gamma \sqrt{g_{ab}\dot{\gamma}^a\dot{\gamma}^b}\,\textrm{d}t.
\end{equation}
Geodesics are the curves that extremizes the length $s$. They can be computed from the geodesic equation, a non-linear partial differential equation:
\begin{equation}\label{eq:geodesic}
    \ddot{\gamma}^a + \Gamma^a_{bc}\dot{\gamma}^b \dot{\gamma}^c = 0,
\end{equation}
where $\Gamma^a_{bc}$ are the Christoffel symbols.\footnote{Ref. \onlinecite{klingenberg2013course}, p. 79.} The Christoffel symbols are computed from the metric\footnote{Ref. \onlinecite{klingenberg2013course}, p. 61.}:
\begin{equation}\label{eq:Gamma}
    \Gamma^\alpha_{\mu\nu} = \frac{1}{2}g^{\alpha\beta}(g_{\beta\mu,\nu} + g_{\beta\nu,\mu} - g_{\mu\nu,\beta}).
\end{equation}
The symbol $g^{\alpha\beta}$ with up-up indices refers to the inverse of the metric $g_{\alpha\beta}$ written with down-down indices. Lowered indices after a comma refers to partial differentiation with respect to the corresponding coordinate: $g_{\beta\mu,\nu}=\partial_{\nu}g_{\beta\mu}$. These are standard notation in differential geometry.

In general relativity, point particles, with or without mass, travel along geodesics of the given spacetime, a four-dimensional metric space. In thermodynamic geometry however, no mechanism constrains systems to evolve exclusively along geodesics. Nevertheless, the significance of these geodesics in thermodynamic geometry has been noted in several papers\cite{ruppeiner1979thermodynamics,ruppeiner2010thermodynamic,diosi1984metricization,feldmann1985thermodynamic,nulton1985quasistatic,salamon1985length}.

The geodesics of the Ruppeiner metric, being the curves that minimize the average fluctuations between two states\cite{diosi1984metricization}, was utilized by Diósi et al.\cite{diosi1989mapping} to make a covariant rule that distinguishes whether two states are of the same phase or not. This effectively partitions the van der Waals state space into its liquid and gas regions, even above the critical point. Since the separation of phases blurs out above the critical point, it is interesting to investigate whether the Diósi partitioning predicts a physical process when systems cross the Diósi boundaries at the supercritical region. It is sensible to ask if the Diósi boundaries in the supercritical region correspond to anything physically meaningful. In other words, we wish to know if there is any measurable difference between states on opposite sides of a supercritical Diosi boundary. However, as we will see below, the Diosi boundaries do not correspond to an established physical marker that separates the supercritical region, and this we consider to be a major weakness of their partitioning scheme. On the other hand, we have discovered that by using the Ruppeiner-$N$ metric, instead of the usual Ruppeiner-$V$ metric of the Diósi partitioning, one can now generate a geodesic-based boundary that separates the supercritical region and that also happens to correspond to a physical thermodynamic boundary. Remarkably, the boundary based on Ruppeiner-$N$ geodesic partitioning is coincident with the isotherm thermodynamic Widom line.

In what follows, we first review the Diósi partitioning based on Ref. \onlinecite{diosi1989mapping} before discussing our own partitioning of the van der Waals state space using the Ruppeiner-$N$ metric.

\subsection{Diósi partitioning of the van der Waals state space}

Just in this subsection, we use the same definition of dimensionless variables used in the paper of Diósi et al. so that comparison can readily be done. The variables are nondimensionalized but \textit{not} normalized, i.e., Eqs. \eqref{eq:crit_1}-\eqref{eq:crit_3} but without the coefficients $1/27$, $8/27$, and $1/3$, respectively. These numbers become the critical values of the dimensionless pressure, temperature, and density in this definition, respectively.

\begin{figure}[h!]
\includegraphics[width=0.5\textwidth]{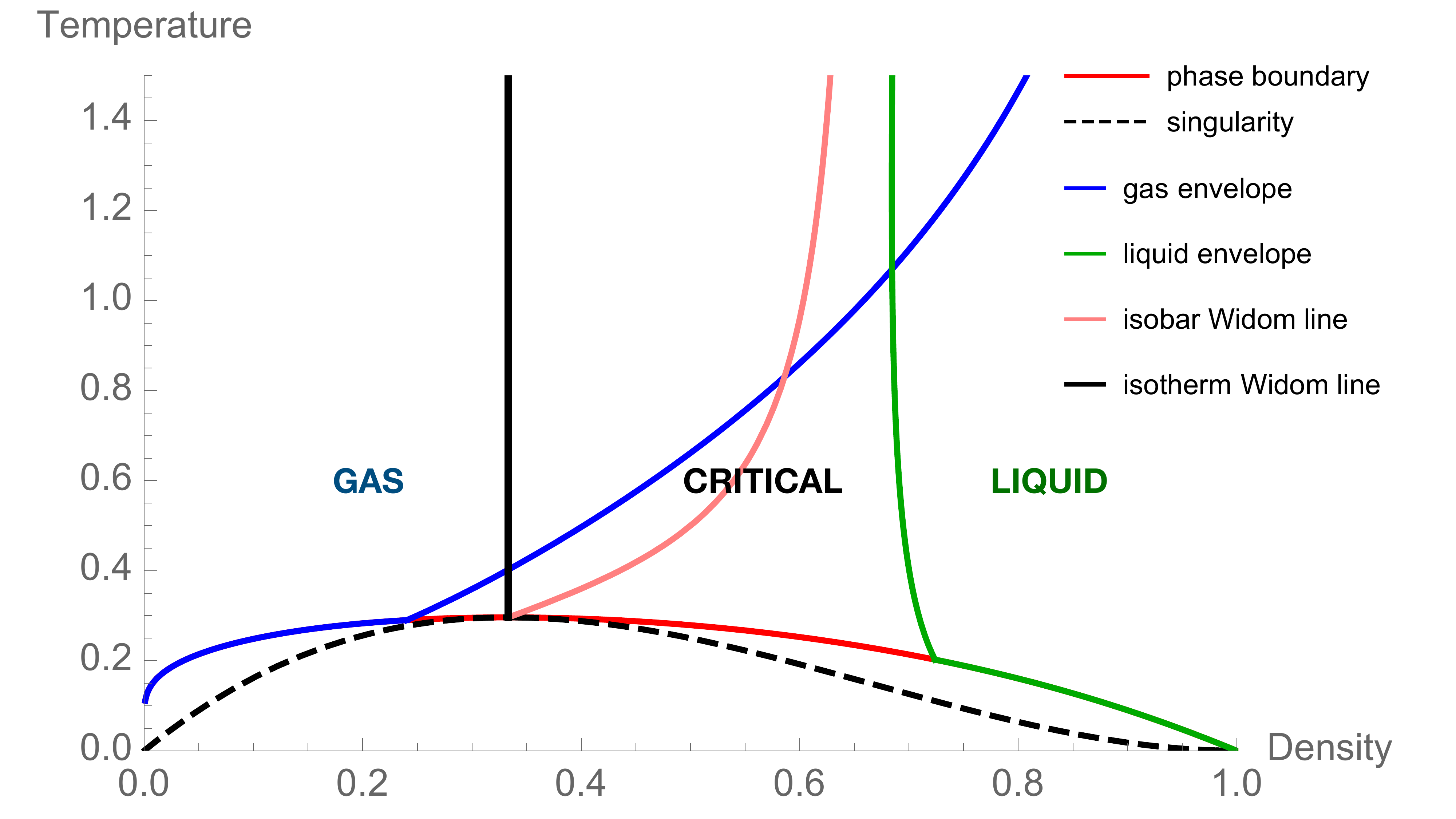}
\caption{\label{fig:diosi} Diósi partitioning of the van der Waals state space.}
\end{figure}

The Diósi scheme for partitioning the van der Waals state space can be succinctly stated as follows:
\begin{quote}
    \textit{Gas (liquid) states are states that are geodesically connected to every gas (liquid) state on the phase boundary}.
\end{quote}
The resulting boundaries are given in Fig. \ref{fig:diosi}. States on the left of the blue curve are the ``gas'' states and states on the right of the green curve are the ``liquid states''. The phase boundary is the red curve and part of it overlaps with the blue and green curve. The Diósi gas and liquid regions intersect somewhere on the supercritical region. States here are both geodesically connected to every Diósi gas and liquid states on the phase boundary. Furthermore, there is a third region bounded by these two Diósi curves, the \textit{critical} region where states are not  geodesically connected to any Diósi gas and liquid states on the phase boundary. Meanwhile, the dashed line is the spinodal curve. It coincides exactly with the boundary of the thermodynamic manifold where the Ricci scalar (as defined in Eq.~\eqref{eq:ricciscalar} below) diverges. The critical point is the peak of the spinodal curve, as seen in the figure.

To generate these boundaries (blue and green curves), we propagated geodesics starting from a sample point at both asymptotic ends of the phase boundary. For example, the (blue) gas boundary is the envelope of geodesics emanating from the asymptotically cold gas state $(x=0.01,T=0.16)$ and above the phase boundary. Meanwhile, the (green) liquid boundary's source point of geodesics is at $(x=0.98,T=0.0196)$. To differentiate this geodesic-based classification of phases from the standard thermodynamic one, we call the former ones the \textit{Diósi gas} and the \textit{Diósi liquid}.

One weakness of the Diósi partitioning that can be readily seen from Fig. \ref{fig:diosi} is that the Diósi boundaries do not neatly slice the phase boundary into the two phases of the van der Waals fluid. This is illustrated in Fig. \ref{fig:diosi}. States lying on the phase boundary with densities lower than the critical density ($x=1/3$) are not geodesically connected to each other (in Fig. \ref{fig:diosi}, the phase boundary are not all colored blue at the left of $x=1/3$). The same can be said about some states on the liquid side. The Diósi partitioning does not even completely coincide with the already established (i.e. usual) thermodynamic partitioning of subcritical states.

Now to test whether the Diósi boundaries correspond to possible physical boundaries at the supercritical region, we superimposed the isotherm (solid black curve) and isobar (orange curve) thermodynamic Widom line atop the Diósi boundaries (see Fig. \ref{fig:diosi}). A sketch for finding the Widom line of a system is given in Subsection V-A. We used Widom lines because these are considered by many authors to be the continuation of the phase boundary on the supercritical region. As can be readily seen, the Diósi boundaries ignore the presence of any of these lines. They fail to connect to any physical features of the van der Waals fluid.

\subsection{Ruppeiner-$N$-based partitioning of van der Waals}

In this subsection, we revert back to the dimensionless and normalized variables in Eqs. \eqref{eq:crit_1}-\eqref{eq:crit_3}.

\begin{figure}[h!]
\includegraphics[width=0.5\textwidth,trim={3cm 1cm 3cm 0},clip]{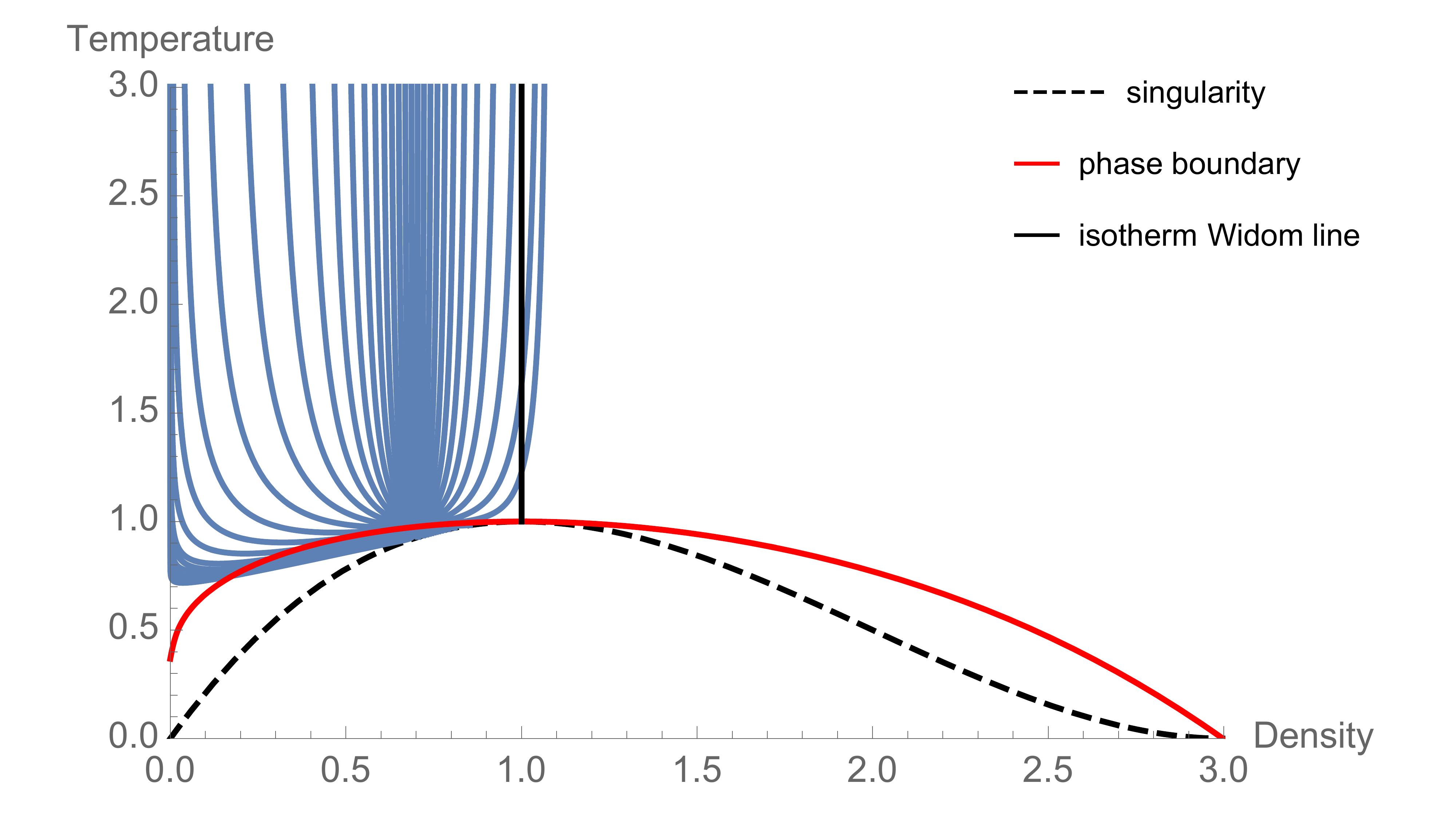}
\caption{\label{fig:diosi-n_1} Geodesics propagating at a gas state on the phase boundary in the Ruppeiner-$N$ van der Waals space. The source point is at $(x=0.7000,T=0.9757)$. Notice that only a few rays cross the critical isochore, which is also the isotherm thermodynamic Widom line of the van der Waals fluid.}
\end{figure}

\begin{figure}[h!]
\includegraphics[width=0.5\textwidth,trim={3cm 1cm 3cm 0},clip]{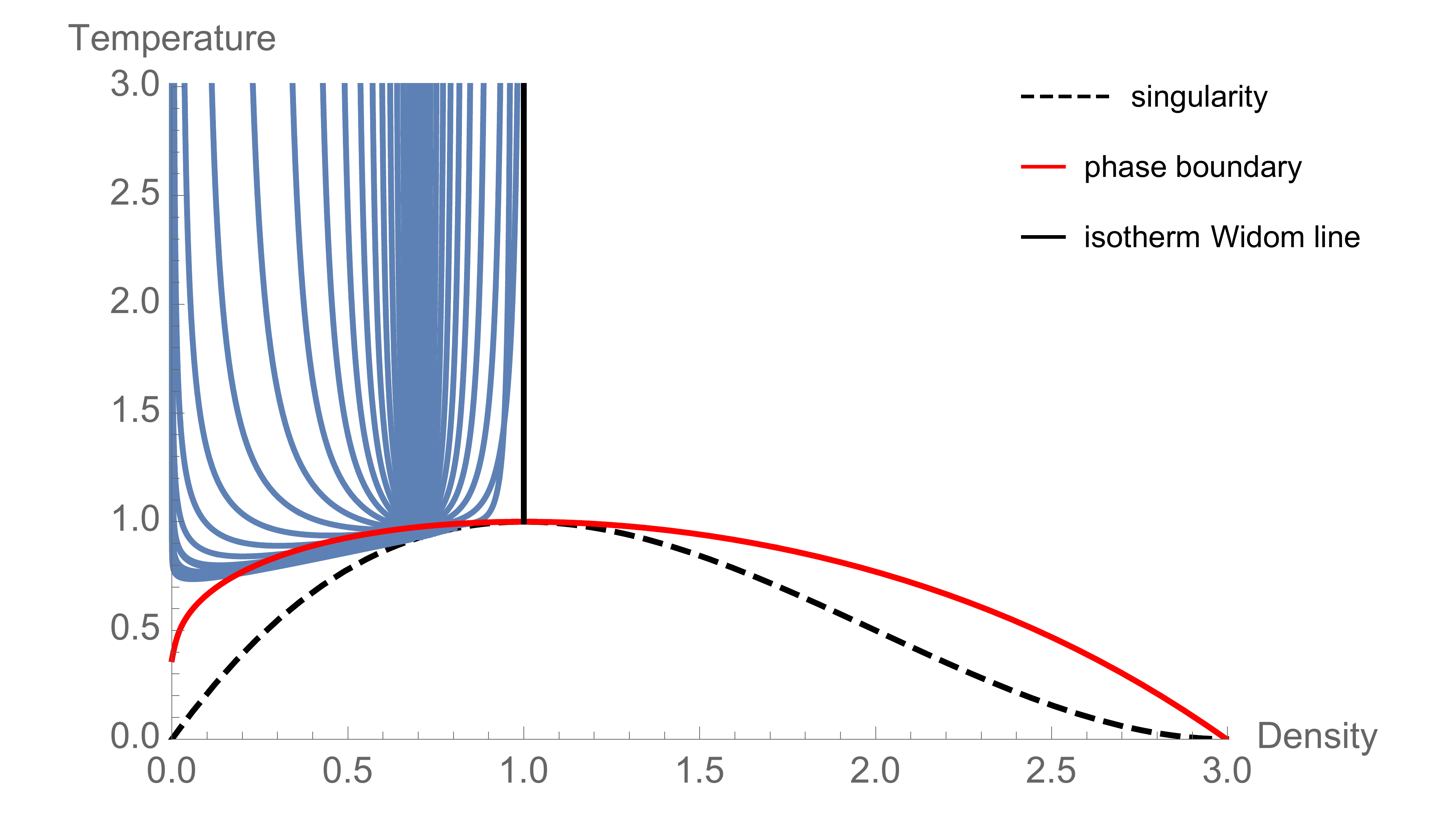}
\caption{\label{fig:diosi-n_2} Geodesics propagating from a gas state slightly below the phase boundary. The source point is at $(x=0.7000,T=0.9657)$, a difference of $-0.01$ from the temperature value of the source point in Fig. \ref{fig:diosi-n_1}. Here, rays are no longer seen to intersect the critical isochore/isotherm Widom line.}
\end{figure}

\begin{figure}[h!]
\includegraphics[width=0.5\textwidth,trim={3cm 1cm 3cm 0},clip]{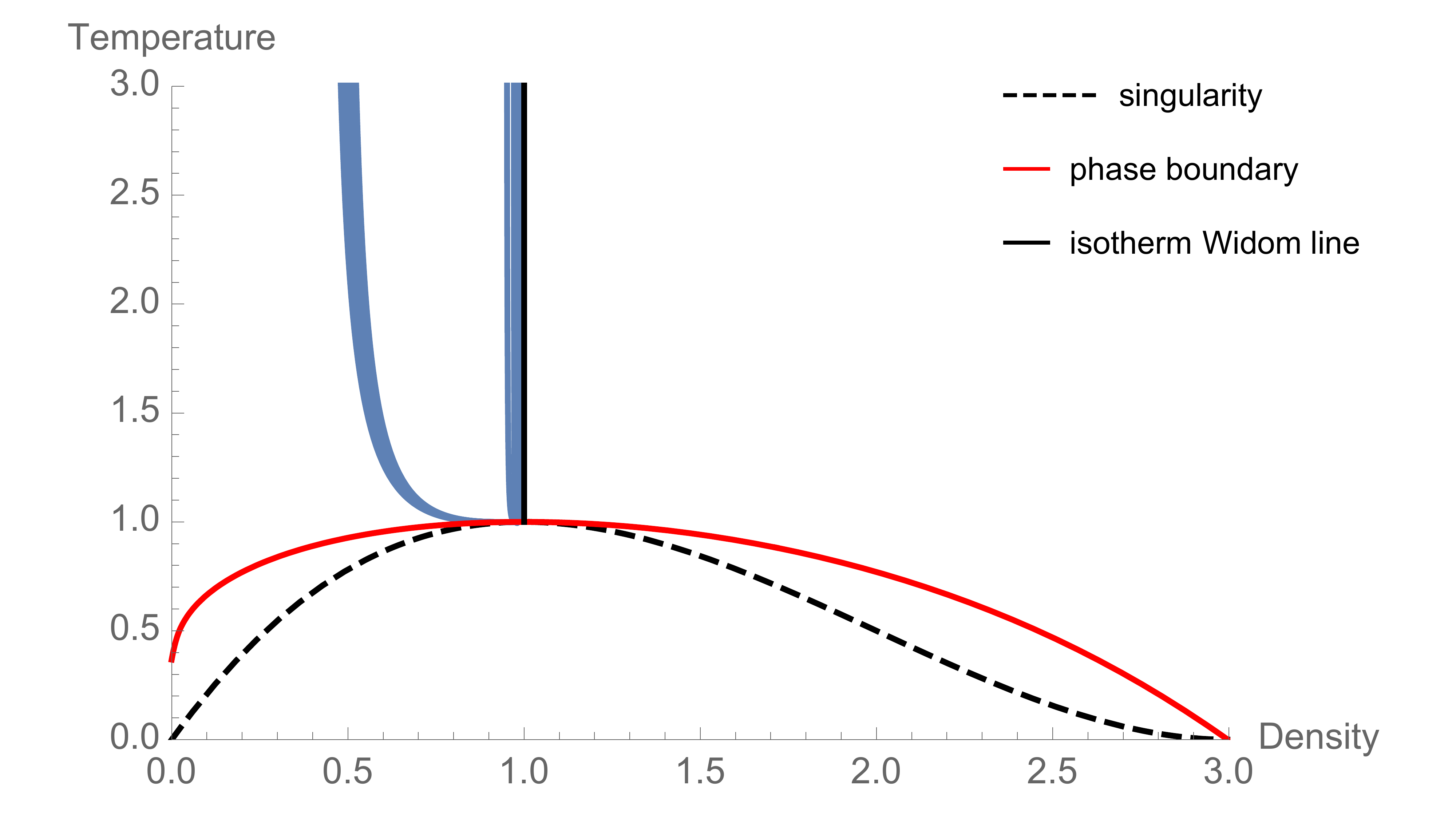}
\caption{\label{fig:diosi-n_3} Geodesics propagating from an asymptotically spinodal gas state very near the critical point. One thousand rays are propagated from the state at $(x=0.999000,T=0.999999)$.}
\end{figure}

The general feature of geodesics using the Ruppeiner-$N$ metric is that geodesics emanating from gas states $(x<1)$ hardly cross the critical isochore $x=1$ with the exception of few trajectories. For example in Fig. \ref{fig:diosi-n_1}, rays at different initial angles in the $(x,T)$ space are propagated from a single point. The chosen initial point is at $(x=0.7000,T=0.9757)$, a gas state on the phase boundary. We can see that only a few rays passes the critical isochore with much of them staying at the gas side (to the left of the critical isochore) of the state space. If we lower down the temperature, fewer rays cross the critical isochore (see Fig. \ref{fig:diosi-n_2}). This prompts us to check the geodesics emanating from asymptotic states just above the spinodal curve. From Fig.~\ref{fig:diosi-n_3}, we can see that the geodesics are compressed and bundled along a vertical line (constant $x$) even though initial angles are varying. Looking at an asymptotic spinodal gas state very near the critical point, we see that no geodesic emanating from this state crosses the critical isochore. That is, all geodesics stay at the left of the critical isochore. The same features hold for the liquid states at the right of the critical isochore. Thus, the geodesics in the Ruppeiner-$N$ space are aware of the presence of the critical isochore. Interestingly, this critical isochore is also the isotherm thermodynamic Widom line of the van der Waals fluid.
We now state our own partitioning rule:
\begin{quote}
    \textit{Gas (liquid) states are states that are geodesically connected to any asymptotic spinodal gas (liquid) state.}
\end{quote}
Our rule generates only one boundary, which is exactly the critical isochore and the isotherm thermodynamic Widom line. Any state on the left of the critical isochore (in Figs. \ref{fig:diosi-n_1}-\ref{fig:diosi-n_3}) is a gas and any state on the right is a liquid. This is entirely consistent with the standard definition of gas and liquid states of the van der Waals fluid on the subcritical region. Furthermore, our boundary coincides with the isotherm Widom line\cite{brazhkin2011van}, a proposed continuation of the phase boundary above the critical point.

\section{Expansion of equation of state near the critical point\label{sec:expand}}

Applications of the Ruppeiner metric and the $R$-crossing method to \emph{specific} model systems are abundant in the literature. In this paper, we wish derive general facts from these applications. For this reason, we construct a fairly general EoS that we can directly insert into the Ruppeiner metric.

We start with a pressure EoS written as a function of temperature and number density $P(T,x)$ and perform an expansion in the number density about the critical point $x=1$. The reason for this expansion is because the $R$-crossing method is expected to work better as one goes near the critical point. This is due to Ruppeiner's hypothesis of the Ricci scalar being proportional to the correlation length near the critical point.\cite{ruppeiner2012thermodynamic} Because at the critical point $\partial_x P = \partial_x^2 P =0$, we keep terms in our expansion up to third order in $x$:
\begin{equation}
    \label{eq:expansion}
    P(T,x) = a_0(T)+a_1(T)x+a_1(T)x^2+a_3(T)x^3.
\end{equation}
We subject Eq. \eqref{eq:expansion} to a number of constrains that limit the forms of the $a_i$ functions so that the pressure EoS evolves as expected as the temperature approaches the critical value $T=1$.

First, we require that below the critical temperature $P(T,x)$ should have two real \textit{extremal} critical points (values of $x$ that extremizes $P$, not to be confused with the thermodynamic critical point). Let $x_1(T)$ and $x_2(T)$ be the extremal critical points of $P$ at a given temperature. Then, we should have
\begin{equation}
    \partial_x P \propto \big(x-x_1(T)\big)\big(x-x_2(T)\big).
    \label{eq:roots_1}
\end{equation}
At the critical temperature, $x_1$ and $x_2$ should coincide at the critical point: $x_1(T=1)=x_2(T=1)=1$. Just above the critical temperature, no extremal critical points should exist because the pressure EoS should now be stable everywhere. So, in general $x_1$ and $x_2$ are complex numbers. We decompose them to their real and imaginary parts
\begin{align}
    x_1(T) &= f_1(T) + i\,g_1(T),\\
    x_2(T) &= f_2(T) + i\,g_2(T).
\end{align}
Even though the $x_1(T)$ and $x_2(T)$ are complex-valued functions, we should not see any imaginary numbers in the full form of $P(T,x)$. Integrating Eq. \eqref{eq:roots_1} with respect to $x$ and zeroing out all imaginary terms, we get the following constraints:
\begin{equation}
   f_1(T) = f_2(T),
\end{equation}
\begin{equation}
   g_1(T) = -g_2(T).
\end{equation}
We now write the extremal critical points as
\begin{equation}
    x_\pm = f(T) \pm i\,\tilde{g}(T).\label{eq:roots_2}
\end{equation}
Now that the behavior of the pressure EoS above the critical point has been worked out, we return to the subcritical region. When $T<1$, $x_\pm(T)$ should be purely real so that the two extremal critical points exist. If we let $\tilde{g}(T)=\sqrt{g(T)}$ where $g(T)$ is a function such that $g(T<1)\leq 0$, $g(T=1)=0$, and $g(T>1)<0$, we see that $x_\pm(T)$ acquires the desired properties we have for it: $x_\pm(T<1)\in \mathbb{R}$, $x_\pm(T=1)=1$, and $x_\pm(T>1) \notin \mathbb{R}$. With this, the final form of our general cubic EoS becomes
\begin{equation}
    P = h(T)\bigg(\frac{1}{3}x^3 - f(T)x^2\ +\Big(f^2(T)-g(T)\Big)x + c(T)\bigg).
    \label{eq:Pform}
\end{equation}
So far, we have four free functions of temperature characterizing a specific system near a critical point. We further impose these constraints to the free functions:
\begin{equation}
    \label{eq:constraint_1}
    P(1,1)=1,
\end{equation}
\begin{equation}
    \label{eq:constraint_2}
    \frac{\text{d}}{\text{d}T}P\big(T,x_\pm(T))\Big|_{T=1}>0,
\end{equation}
\begin{equation}
    \label{eq:constraint_3}
    \frac{\partial P}{\partial x} \Big|_{T>1}>0.
\end{equation}
Equation \eqref{eq:constraint_1} sets the critical value of the dimensionless pressure to unity. Equation \eqref{eq:constraint_2} ensures that the equation $P=P(T<1,x)$ will not have multiple roots in $x$ for $P>1$. This removes the possibility of having coexistent states with pressures greater than the critical value. Finally, Eq. \eqref{eq:constraint_3} imposes that the system be stable above the critical temperature.

The corresponding Helmholtz free energy is
\begin{equation}
    F = h(T)\bigg(\frac{1}{6}x^2 - f(T)x\ +\Big(f^2(T)-g(T)\Big)\ln{x} - \frac{c(T)}{x}\bigg) + j(T).
    \label{eq:Fform}
\end{equation}
calculated from the relation\footnote{Ref. \onlinecite{callen1985thermodynamics}, Section 5-2.} $P = (\partial G/ \partial V )_T$. Notice the additional free function $j(T)$.

\subsection*{Choice of free functions}

We have a total of five free functions in our expansion, $c(T)$, $f(T)$, $g(T)$, $h(T)$, and $j(T)$, that characterize a thermodynamic system near a critical point. For example, the van der Waals fluid with the pressure\footnote{Ref. \onlinecite{gould2010statistical}, p. 371.} and temperature\footnote{Ref. \onlinecite{callen1985thermodynamics}, p. 76.} EoS
\begin{equation}
    P= \frac{8xT}{3-x} - 3x^2\label{eq:full_VDW_P_EoS}
\end{equation}
\begin{equation}
    T= \frac{1}{4}(E+3x)\label{eq:full_VDW_T_EoS}
\end{equation}
using normalized variables ($E$ stands for the normalized internal energy) has the following assignment in our expansion:
\begin{align}
    c(T)&=-1/9,\label{eq:una}\\
    f(T)&=1/3\,+\,2/(3T),\\
    g(T)&=f^2(T)-1,\\
    h(T)&= 9T/2,\\
    j(T)&=-4T\big(\ln{(T)} + K\big),\label{eq:dulo}
\end{align}
where $K$ is a constant. 

In this paper, we consider three other choices of free functions that correspond to three toy systems. We name these as Systems Maria, Rose, and Sina.\footnote{From Maria, Rose, and Sina who acquired the Power of the Titans from their mother Ymir in the manga and anime adaptation \textit{Attack on Titan}.} These systems are not necessarily physically realizable, but we want to examine how the Ruppeiner metric behaves under a bare EoS satisfying minimal properties, in this case one that is analytic and features a critical point. System Maria has
\begin{align}
    c(T)&=2/3-(1-T),\\
    f(T)&=1,\\
    g(T)&=1-T,\\
    h(T)&=1,\\
    j(T)&=-T^2/2;
\end{align}
System Rose has
\begin{align}
    c(T)&=2/3-(1-T),\\
    f(T)&=1,\\
    g(T)&=(1-T)+(1-T)^2,\\
    h(T)&=1,\\
    j(T)&=-T^2/2;
\end{align}
and System Sina has
\begin{align}
    c(T)&=2/3-(1-T),\\
    f(T)&=T,\\
    g(T)&=1-T,\\
    h(T)&=1,\\
    j(T)&=-T^2/2.
\end{align}

The three toy systems differ mainly in their definition of the functions $f(T)$ and $g(T)$, the functions that model the spinodal curve
\begin{equation}
    x_\pm(T) = f(T) \pm \sqrt{g(T)},
\end{equation}
as can be seen in Eq. \eqref{eq:roots_2}. System Maria has the simplest functions: a constant $f(T)$ and a linear $g(T)$. System Rose features an additional quadratic term in $g(T)$ compared to System Maria. Lastly, System Rose is System Maria with a linear $f(T)$. Meanwhile, the function $j(T)$ sets the isochoric heat capacity of the system from the relation $C_V = -T(\partial^2 F/\partial T^2)_V$.\cite{kumar2012geodesics}

\section{\label{sec:ricci}Ricci construction of the phase boundary and the isobar Widom line}

In this section, we discuss the applications of Ruppeiner's hypothesis in Eq. \eqref{eq:Rpropto} that the Ricci scalar is proportional to the correlation length near the critical point of a system. As discussed in Section \ref{sec:intro}, Ruppeiner's hypothesis introduces an alternative method for computing the phase boundary and the (isobar) Widom line, which we collectively call the \textit{Ricci construction}. The summary of the methods is presented in Table \ref{tab:summary}. We tested the Ricci  construction to the van der Waals fluid and the three model systems we presented in Section \ref{sec:expand}. Correspondingly, we generated the same thermodynamic lines using standard methods in thermodynamics and compared the results. The methods are discussed in the following subsection.
\begin{table}[h!]
\caption{\label{tab:summary}Terminology and definitions of the standard/thermodynamic and Ricci-constructed phase boundary and isobar Widom line.}
\begin{ruledtabular}
\begin{tabular}{ >{\raggedright\arraybackslash}p{0.2\linewidth}  >{\raggedright\arraybackslash}p{0.4\linewidth} >{\raggedright\arraybackslash}p{0.4\linewidth} }
&standard construction&Ricci construction\\
\hline
phase boundary & Maxwell phase boundary\newline(coexistent states have equal Gibbs free energy) & Ricci phase 
boundary\newline(coexistent states have equal Ricci scalar)\\
\hline
isobar Widom line & Widom line\newline(maximum isobaric heat capacity along isobars) & Ricci-Widom line\newline(maximum Ricci scalar along isobars)\\
\end{tabular}
\end{ruledtabular}
\end{table}

\subsection{\label{subsec:metho}Methodology}

\subsubsection*{Standard thermodynamic construction}

The phase boundary and the (thermodynamic) Widom line can be readily computed using standard methods in thermodynamics. The standard Maxwell phase boundary is computed by finding pairs of states with different densities, $x_A$ and $x_B$, for every value of the temperature less than the critical value that satisfies the equality of pressure as dictated by a pressure EoS
\begin{equation}
    \label{eq:p_eos_equal}
    P(T,x_A)=P(T,x_B),
\end{equation}
and the equality of the Gibbs free function (thus, the chemical potential $\mu=G/N$ as well)\footnote{Ref. \onlinecite{callen1985thermodynamics}, p. 221.}
\begin{equation}
    \label{eq:G_equal}
    G(T,P,x_A)=G(T,P,x_B).
\end{equation}
The two unknowns $x_A$ and $x_B$ are completely determined from the two equations Eqs. \eqref{eq:p_eos_equal} and \eqref{eq:G_equal}.

Meanwhile, the isobar thermodynamic Widom line is directly computed by finding the maximum of the isobaric heat capacity at lines of constant pressure. We use the isobar variant of the Widom line as this is the definition most often used in the literature\cite{lamorgese2018widom,ruppeiner2012thermodynamic,banuti2015crossing}. Given a pressure EoS $P(T,V,N)$ and a corresponding Helmholtz free energy $F(T,V,N)$, the isobaric heat capacity can be calculated using
\begin{equation}
    C_P= -T\Big(\frac{\partial^2 F}{\partial T^2}\Big)_V -T\Big(\frac{\partial P}{\partial T}\Big)^2_V \Big(\frac{\partial P}{\partial V}\Big)^{-1}_T.\label{eq:cp}
\end{equation}

\subsubsection*{Ricci construction}

The Ricci scalar $R$ is the simplest curvature invariant that can be obtained from two successive contractions of the Riemann tensor ${R^{\rho}}_{\sigma\mu\nu}$\cite{szekeres2005course}
\begin{equation}\label{eq:riem}
    {R^{\rho}}_{\sigma\mu\nu} = \Gamma^\rho_{\nu\sigma,\mu} - \Gamma^\rho_{\mu\sigma,\nu} + \Gamma^\rho_{\mu\lambda}\Gamma^\lambda_{\nu\sigma} - \Gamma^\rho_{\nu\lambda}\Gamma^\lambda_{\mu\sigma},
\end{equation}
\begin{equation}\label{eq:riccitensor}
    R_{\sigma\nu} = {R^{\mu}}_{\sigma\mu\nu},
\end{equation}
\begin{equation}\label{eq:ricciscalar}
    R = g^{\sigma\nu}R_{\sigma\nu}.
\end{equation}
For diagonal metrics, which are the ones we will work with, the curvature is given by\footnote{For example, Ref. \onlinecite{ruppeiner2010thermodynamic}.}
\begin{equation}
    R = \frac{1}{\sqrt{g}}\bigg[\frac{\partial}{\partial x^1}\Big(\frac{1}{\sqrt{g}}\frac{\partial g_{22}}{\partial x^1}\Big) + \frac{\partial}{\partial x^2}\Big(\frac{1}{\sqrt{g}}\frac{\partial g_{11}}{\partial x^2}\Big)\bigg].
\end{equation}

Widom\cite{widom1974critical} argued that the correlation length must be equal for coexistent states. And by Ruppeiner's hypothesis in Eq. \eqref{eq:Rpropto}, the Ricci scalar must also be equal for two coexistent states. Thus, the Ricci phase boundary is computed from Eq. \eqref{eq:p_eos_equal} and, this time, from the equality of the Ricci scalars
\begin{equation}
    R(T,x_A)=R(T,x_B),
    \label{eq:ricci_equal}
\end{equation}
in place of Eq. \eqref{eq:G_equal}.

Ruppeiner's hypothesis also implies that near the critical point the isobar (isotherm) statistical Widom line should be calculable from the Ricci scalar. Since the Ricci scalar is proportional to the correlation length in this region, one can locate instead the maximum of the Ricci scalar along isobars (isotherms) instead of the correlation length. That is, we collect the values of the number density $x$ that maximize $R(T(P,x),x)$ for a given pressure $P$. This collection of points $(x,P)$ is supposed to be the isobar statistical Widom line. We call this Widom line generated using the Ricci scalar the \textit{Ricci-Widom line}. While the standard construction of the Widom line is thermodynamics-based, the strength of the Ricci-Widom line is that it appeals to the original statistical definition of the Widom line, i.e., the locus of correlation length maxima. In other words, the standard Widom line computed in this paper is an isobar thermodynamic Widom line and the Ricci-Widom line is an isobar statistical Widom line. 

\subsection{\label{sec:ricci_results}Results}

\begin{figure*}
\begin{minipage}{0.49\textwidth}
\includegraphics[width=\linewidth,trim={8cm 1cm 8cm 0},clip]{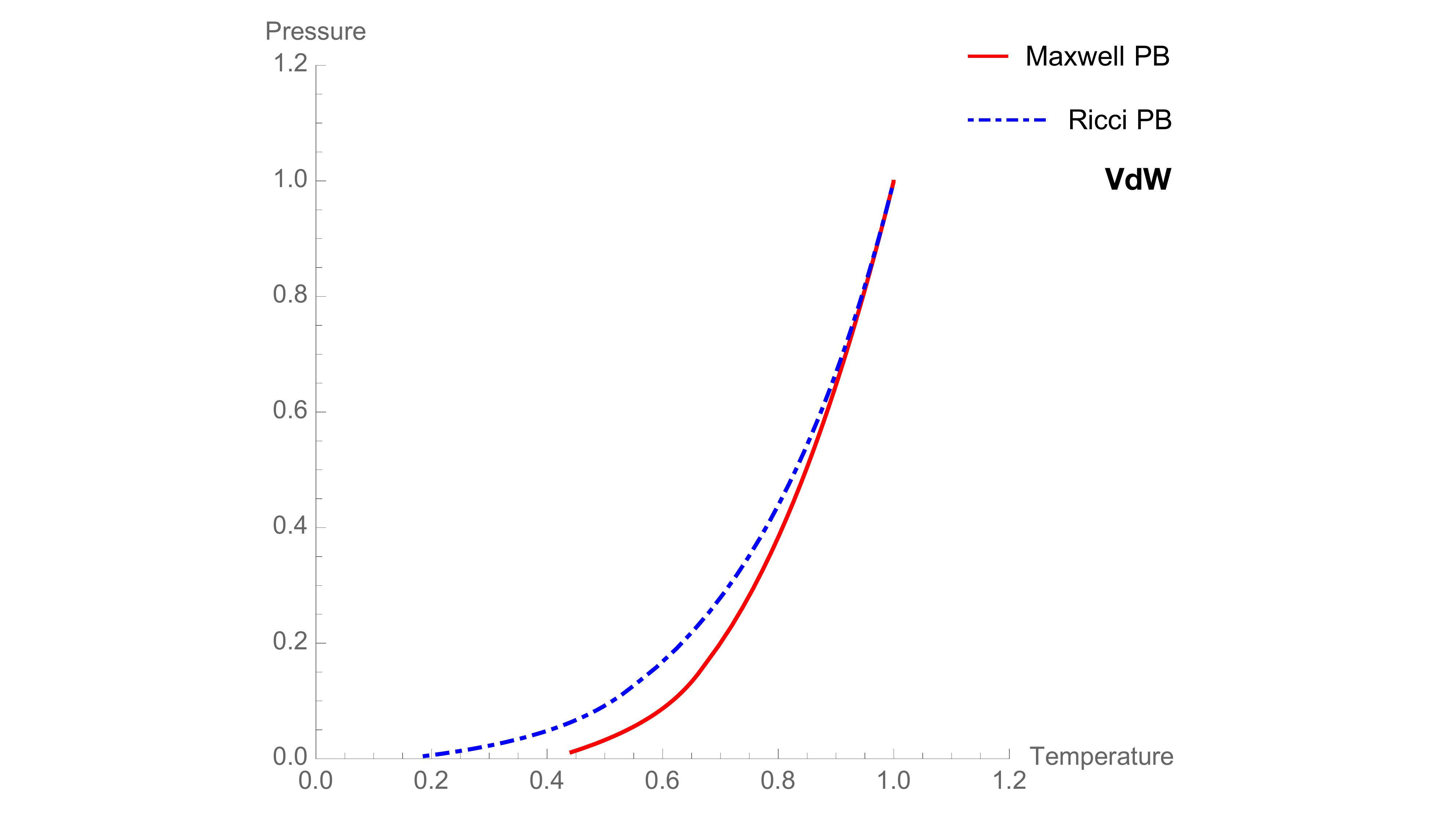}
\end{minipage}
\begin{minipage}{0.499\textwidth}
\includegraphics[width=\linewidth,trim={8cm 1cm 8cm 0},clip]{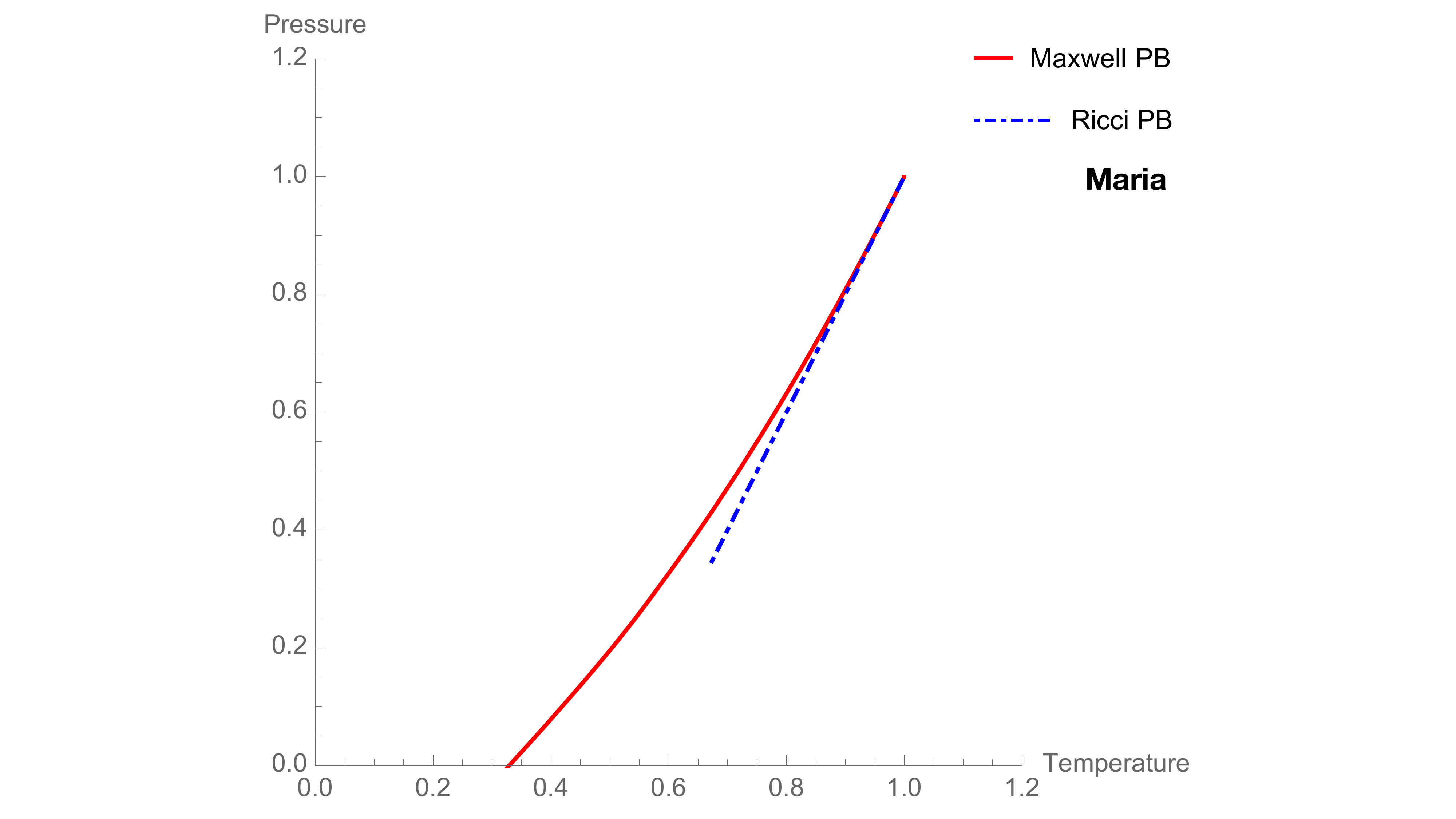}
\end{minipage}
\begin{minipage}{0.499\textwidth}
\includegraphics[width=\linewidth,trim={7.2cm 1cm 7.2cm 0},clip]{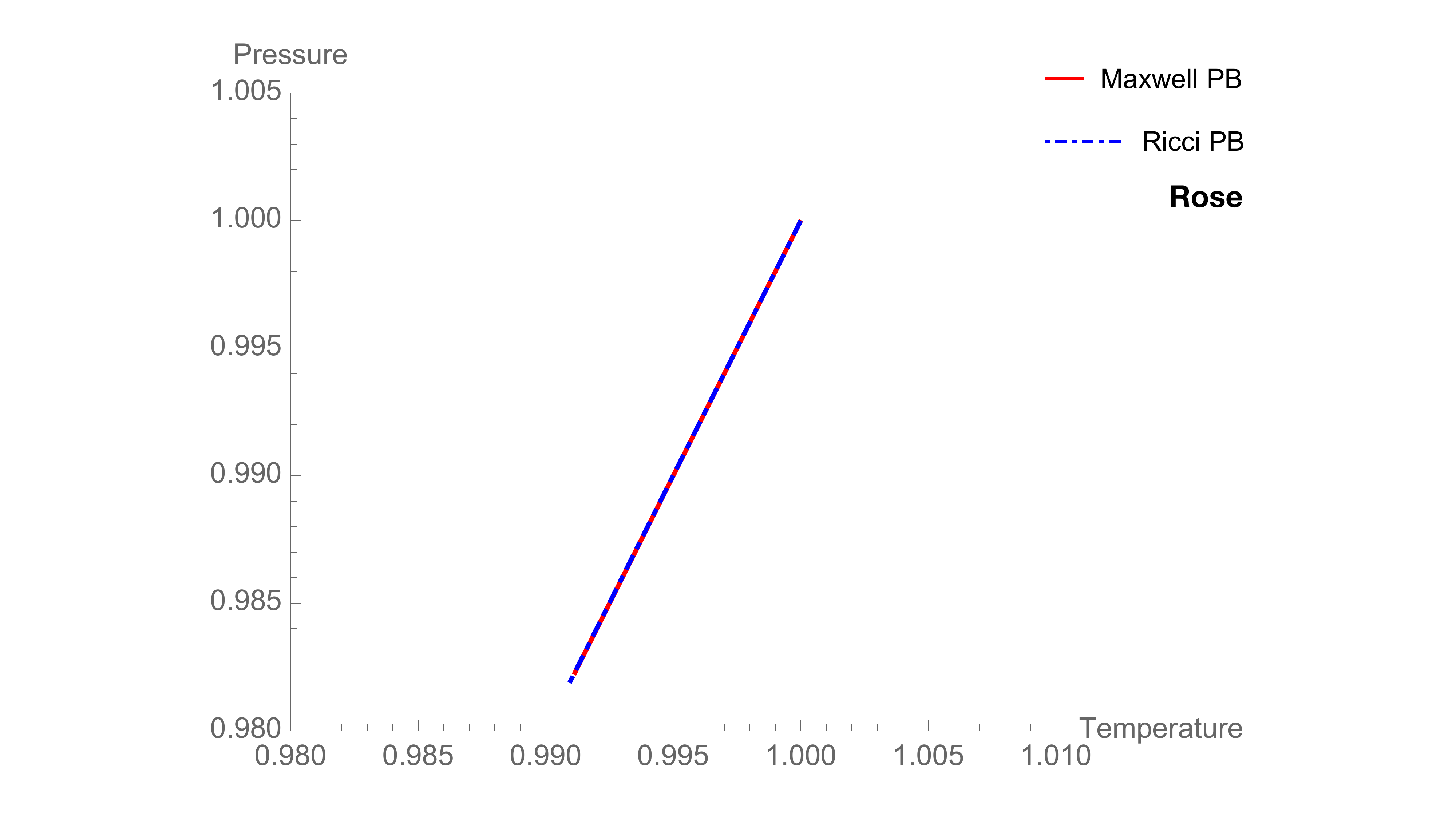}
\end{minipage}
\begin{minipage}{0.49\textwidth}
\includegraphics[width=0.95\linewidth,trim={7.2cm 1cm 7.2cm 0},clip]{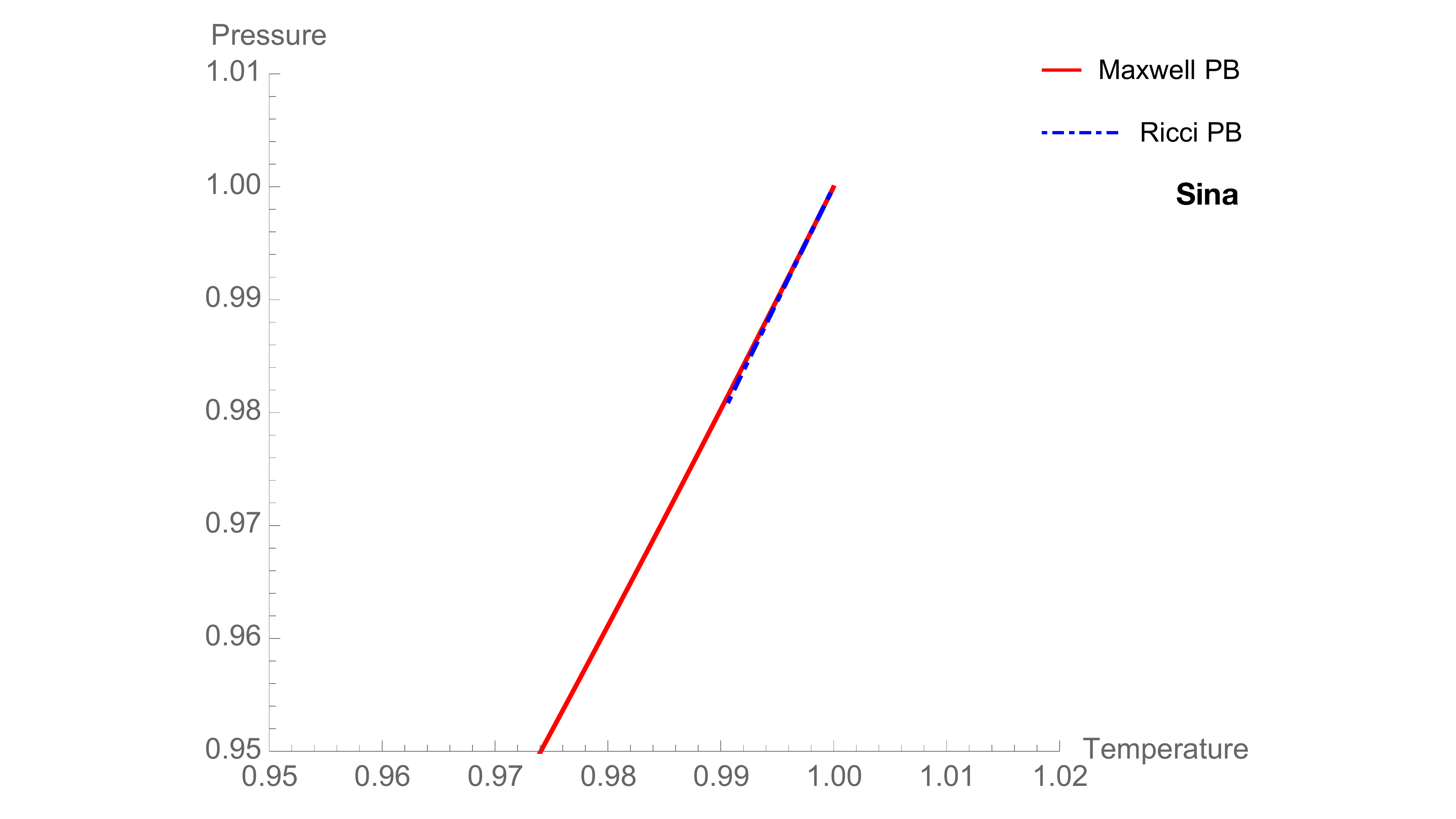}
\end{minipage}
\caption{\label{fig:PT}Maxwell and Ricci phase boundaries of the van der Waals fluid and the toy systems Maria, Rose, and Sina as plotted in a pressure-temperature frame. Normalized thermodynamic variables are used. For the toy systems, physical solutions to Eq. \eqref{eq:ricci_equal} do not exist at temperatures far from the critical value, thus plot ranges are adjusted to zoom in on the curves.}
\end{figure*}

\begin{figure*}
\begin{minipage}{0.49\textwidth}
\includegraphics[width=\linewidth,trim={2.25cm 1cm 5cm 0},clip]{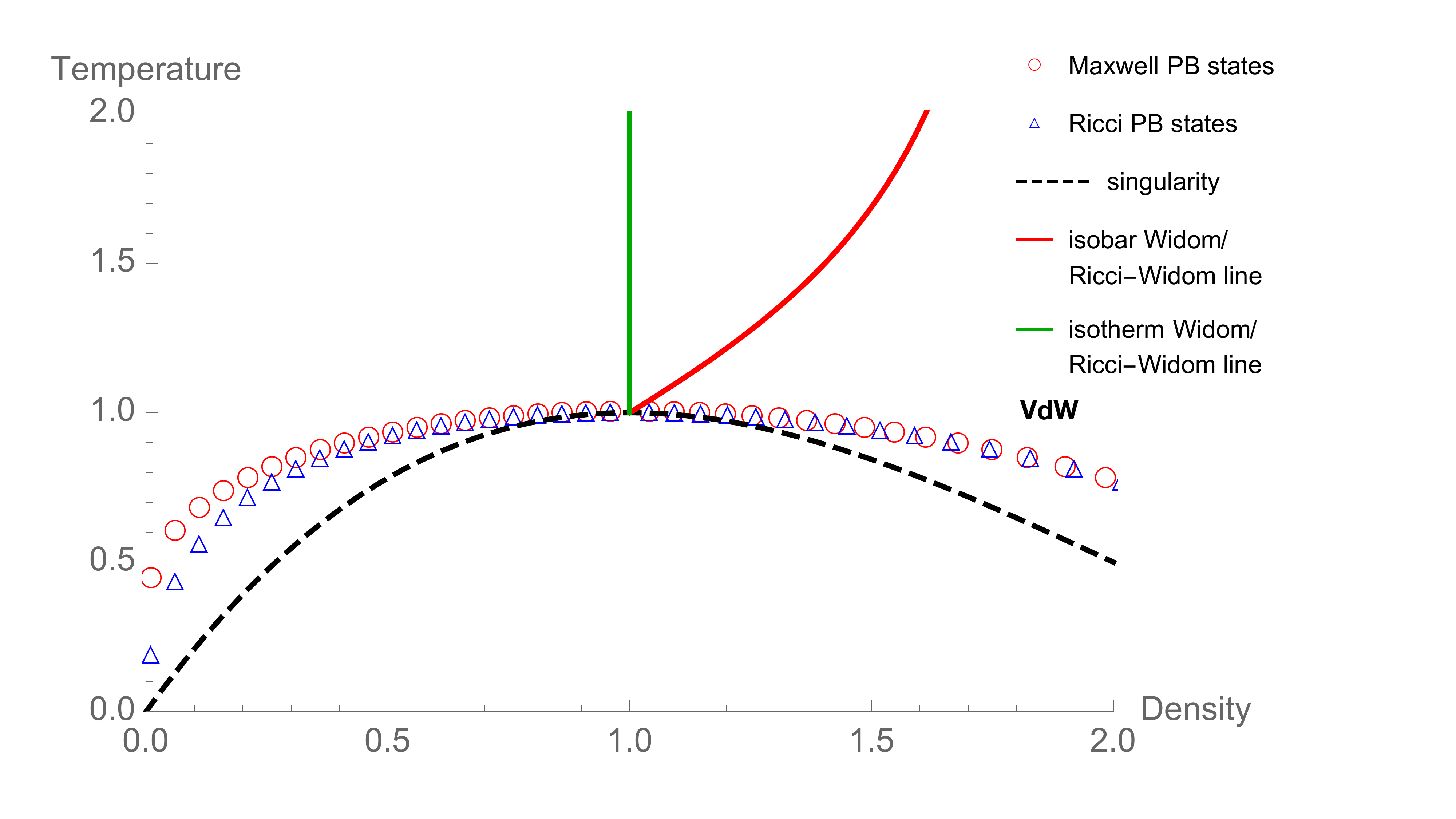}
\end{minipage}
\begin{minipage}{0.49\textwidth}
\includegraphics[width=\linewidth,trim={4cm 1cm 4cm 0},clip]{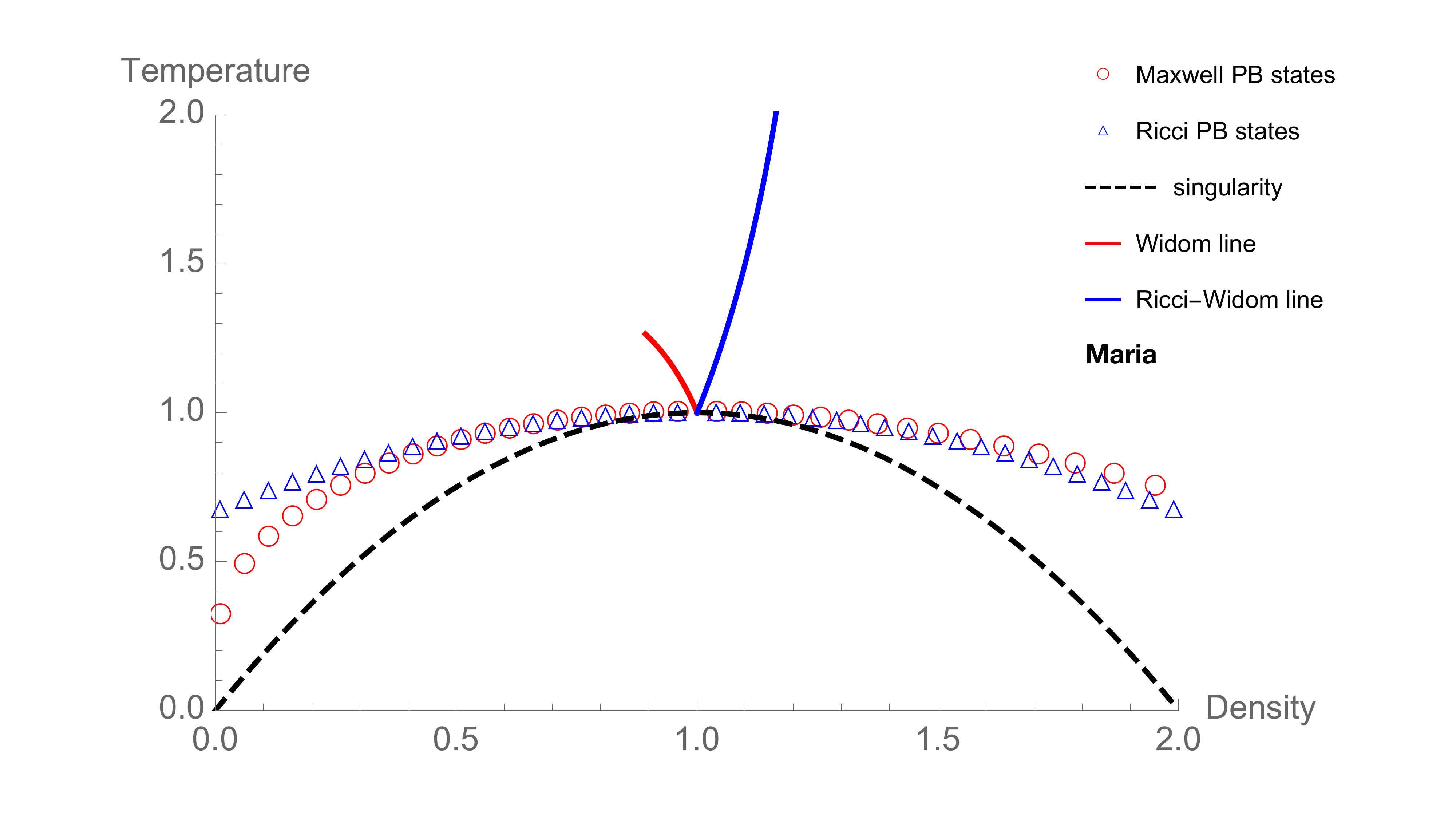}
\end{minipage}
\begin{minipage}{0.49\textwidth}
\includegraphics[width=\linewidth,trim={4cm 1cm 4cm 0},clip]{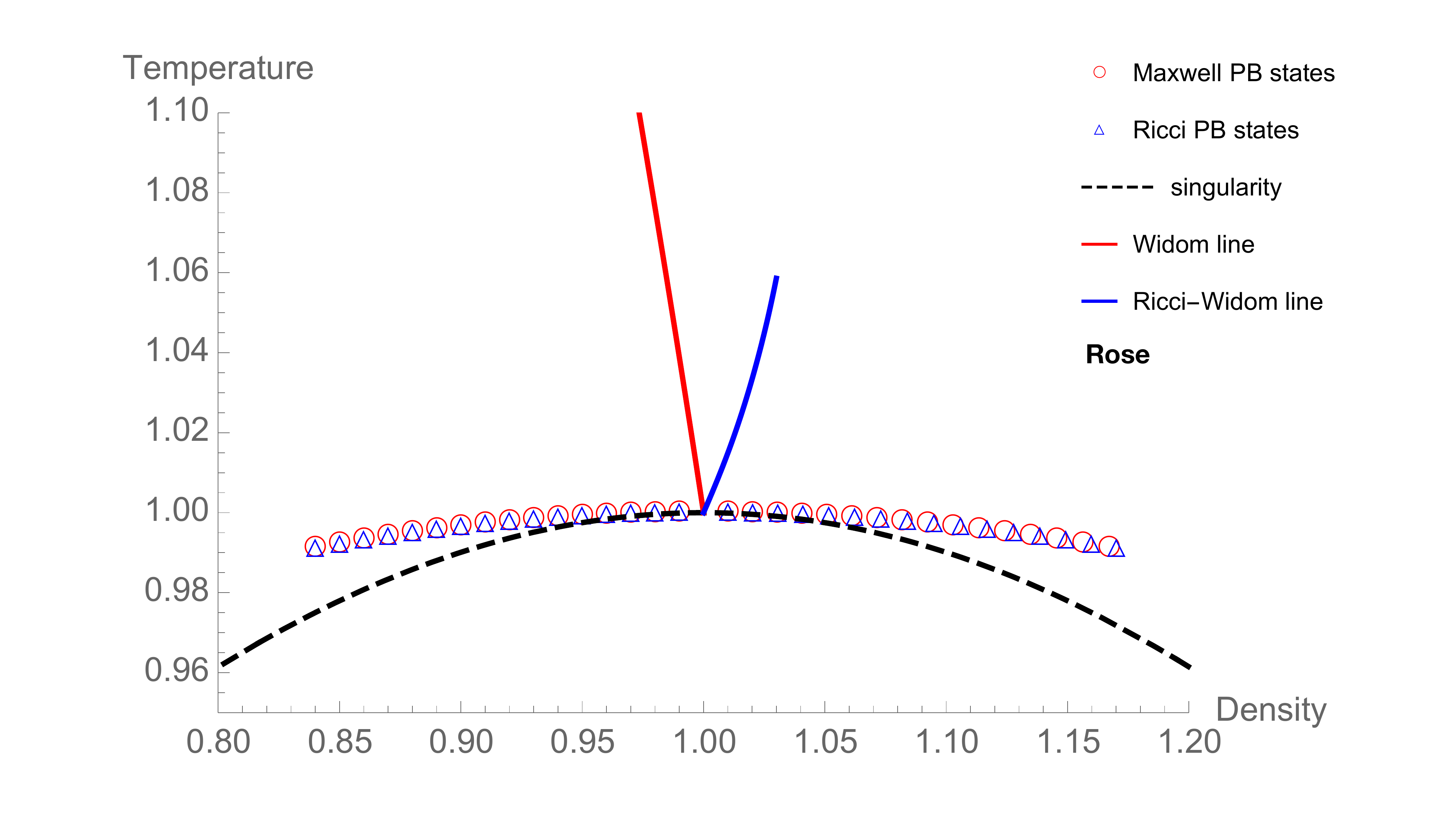}
\end{minipage}
\begin{minipage}{0.49\textwidth}
\includegraphics[width=\linewidth,trim={4cm 1cm 4cm 0},clip]{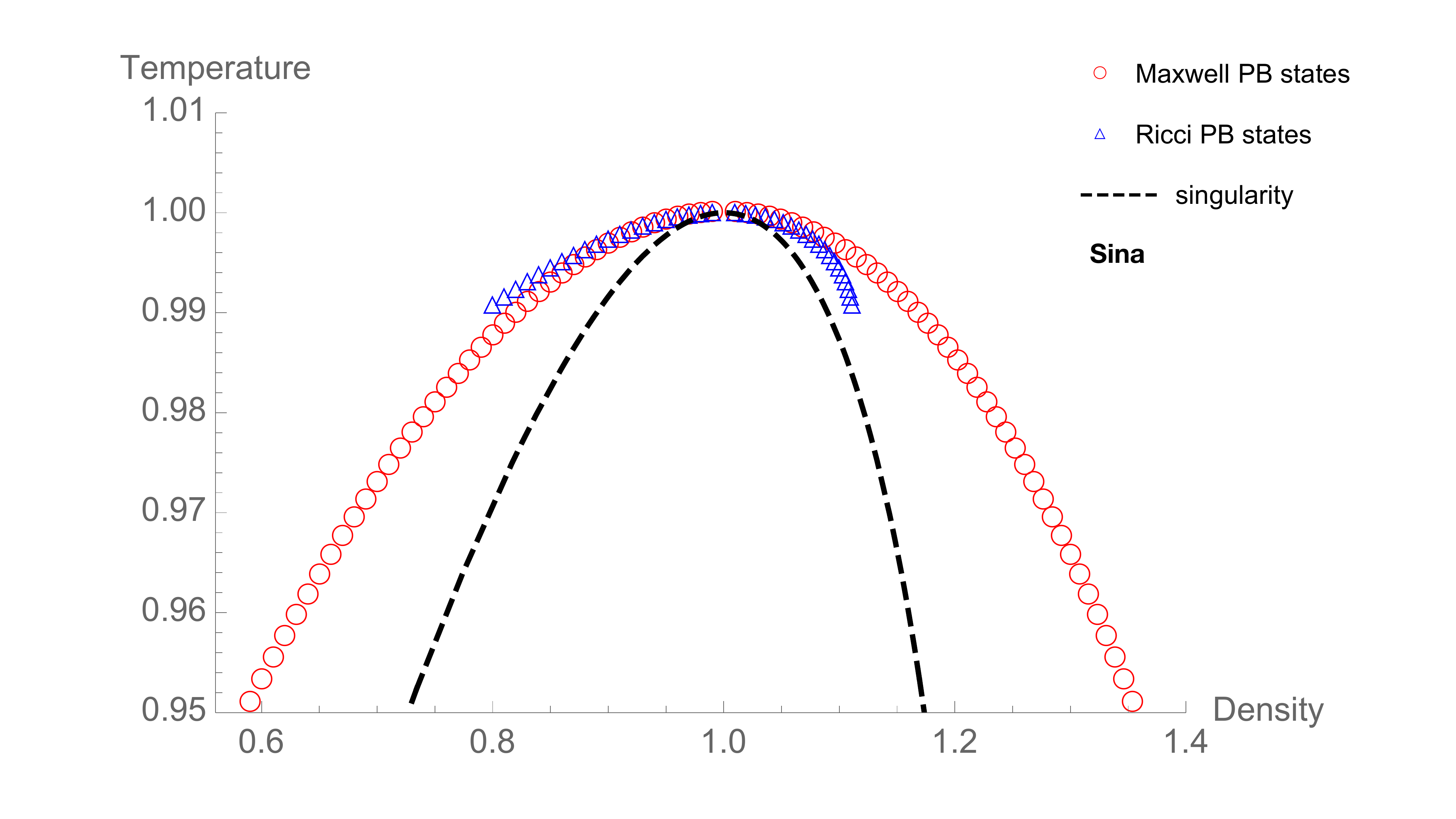}
\end{minipage}
\caption{\label{fig:Tx}Maxwell and Ricci phase boundaries of the van der Waals fluid and the toy systems Maria, Rose, and Sina as plotted in a temperature-density frame. Normalized thermodynamic variables are used. The Ricci-Widom lines of Systems Maria and Rose do not span far enough from the critical point because their Ricci curvatures no longer have relative maxima at pressures far from the critical value. System Sina does not have an isobar thermodynamic Widom line. It has a Ricci-Widom line but this is not shown here shown because it cannot be compared to a standard curve.}
\end{figure*}

The standard and Ricci-based phase boundary and isobar Widom line for the van der Waals fluid and the three toy systems (Systems Maria, Rose, and Sina) are shown in Figs. \ref{fig:PT} and \ref{fig:Tx}. Figure \ref{fig:PT} plots of the phase boundaries in the $P$-$T$ plane while Fig. \ref{fig:Tx} plots the coexistent states and the Widom lines of the sample thermodynamic systems.

\subsubsection*{Proof of equivalence of Maxwell and Ricci phase boundaries}

The general behavior of the Maxwell and Ricci phase boundaries as seen in Figs. \ref{fig:PT} and \ref{fig:Tx} is that they coincide near the critical point. The Maxwell and Ricci phase boundaries are tangent to each other at the critical point. Our plots in Figs. \ref{fig:PT} and \ref{fig:Tx} are qualitatively similar to the calculated Maxwell phase boundary of more practical fluid models\cite{mausbach2019thermodynamic,jaramillo2019r}. It seems then that this agreement of the Maxwell and Ricci phase boundaries near the critical point is a property of a large group of systems, if not universal.

\begin{figure}[H]
    \includegraphics[width=\linewidth,trim={9cm 3.8cm 9cm 5.5cm},clip]{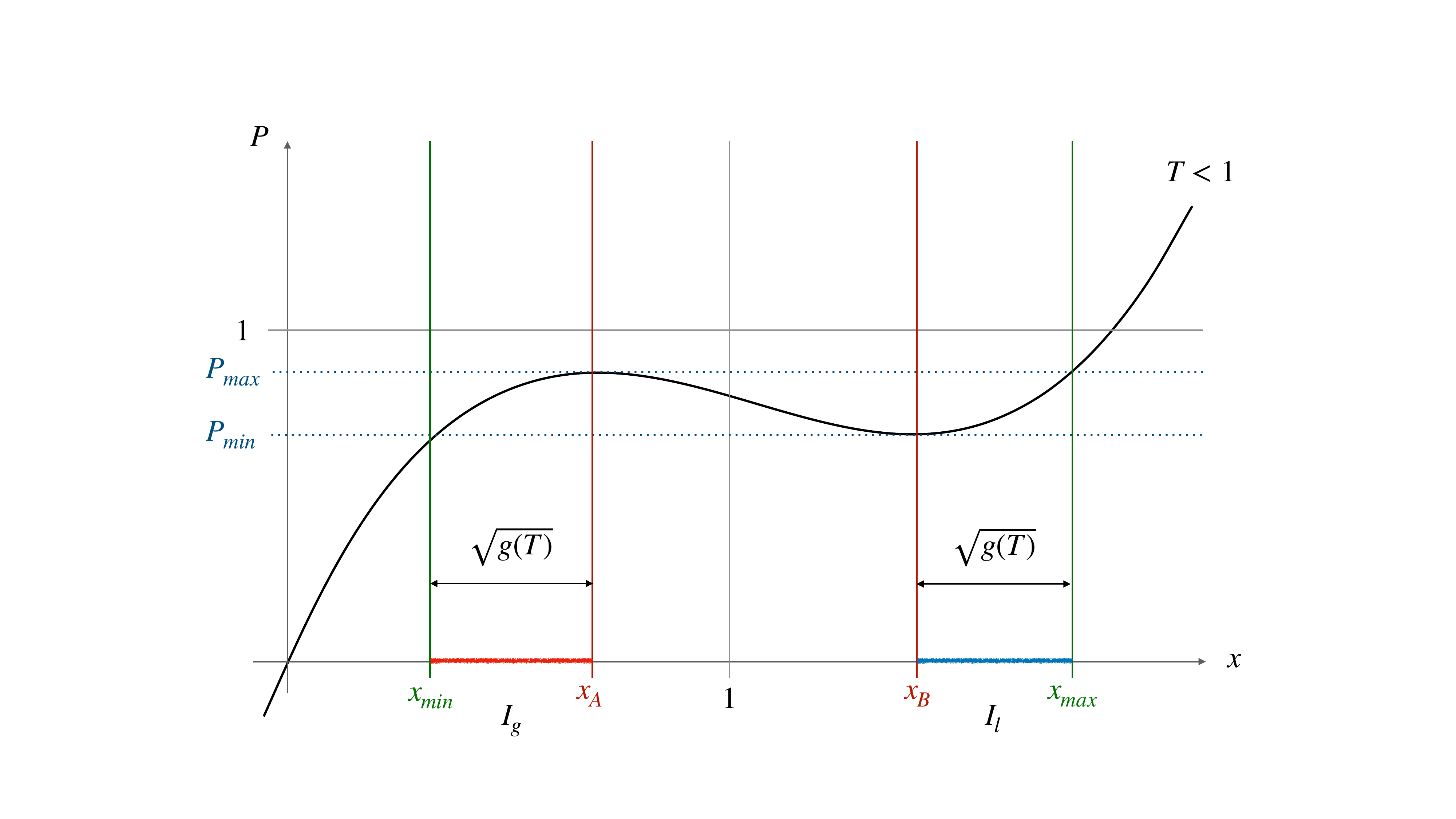}
    \caption{\label{fig:discussion_diagram}A pressure EOS at a temperature below and near the critical value.}
\end{figure}

Here, we prove that the Maxwell and Ricci phase boundaries are guaranteed to agree near the critical point for systems with analytic EoS. Figure \ref{fig:discussion_diagram} shows a generic pressure EoS plot at a subcritical temperature in a pressure-density frame.
The equation $P=P(T,x)$ has multiple density roots in the interval $[P_{min},P_{max}]$. These roots lie in the interval $[x_{min},x_{max}]$. However, roots in the interval $[x_{A},x_{B}]$ correspond to thermodynamically unstable states. The densities $x_A$ and $x_B$ are the densities of the spinodal gas and liquid states, respectively. Stable multiple roots, then, must lie along the intervals $I_g=[x_{min},x_A)$ and $I_l=(x_B,x_{max}]$. This means that coexistent states (roots that additionally satisfy Eq. \eqref{eq:G_equal} for the Maxwell phase boundary states and \eqref{eq:ricci_equal} for the Ricci phase boundary states) must also be in $I_g \cup I_l$. Coexistent gas states are in $I_g$ and coexistent liquid states are in $I_l$. Near the critical point where the EoS may be expressed using our expansion in Eq. \eqref{eq:Pform}, the measure of both intervals is given by $\sqrt{g(T)}$. However, recall that we demanded this free function to vanish at the critical point so that the pressure EoS may feature a critical point. Thus, as the temperature approaches the critical value, the length of the intervals $I_g$ and $I_l$ approaches zero. These narrowing intervals $I_g$ and $I_l$ squeeze the Maxwell and Ricci phase boundaries together until they meet at the critical point.

\begin{figure}[h]
\centering
    \includegraphics[width=\linewidth,trim={1.08cm 0 0 0},clip]{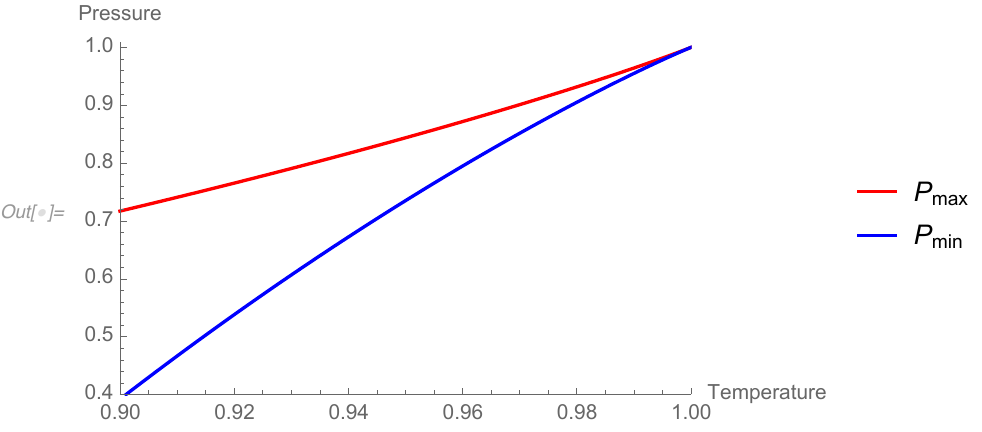}
    \caption{\label{fig:proof}Bounding curves of the phase boundary on the $P$-$T$ plane. This sample plot uses the expanded van der Waals EOS Eqn. \eqref{eq:una}-\eqref{eq:dulo}.}
\end{figure}

Turning to the phase boundary as plotted in the pressure-temperature frame, we can also show that Maxwell and Ricci phase boundaries are tangent to each other at the critical point, as can be seen in Fig. \ref{fig:PT}. The curve $P_{min}$ as function of $T$ is given by $P_{min}=P(T,x_{min}(T))$, and the curve $P_{max}$ as a function of $T$ by $P_{max}=P(T,x_{max}(T))$:
\begin{align}
    P_{min}(T) = \frac{1}{3} h(T) \big(3 c(T)-3 f(T) g(T)+f(T)^3-2 g(T)^{3/2}\big),\label{eq:pmin}\\
    P_{max}(T)= \frac{1}{3} h(T) \big(3 c(T)-3 f(T) g(T)+f(T)^3+2 g(T)^{3/2}\big). \label{eq:pmax}
\end{align}
A sample plot of these curves are shown in Fig. \ref{fig:proof}. Let $P_M(T)$ and $P_R(T)$ be the Maxwell and Ricci phase boundaries in the $P-T$ plane. As discussed previously, $P_{max}(T)$ and $P_{min}(T)$ bounds $P_M(T)$ and $P_R(T)$ from above and below, respectively. We note two things: 1) $P_{max}(T=1)=P_{min}(T=1)$, and 2) $P_{max}'(T=1)=P_{min}'(T=1)$ since
\begin{equation}
    |{P'}_{max}(T)-{P'}_{min}(T)|=\frac{2}{3} \sqrt{g(T)}\, \Big| 3 h(T) g'(T)+2 g(T) h'(T) \Big|\label{eq:lookatthis}
\end{equation}
and $g(T=1)=0$. Because the bounding curves meet at the critical point with the same slope, the Maxwell and Ricci phase boundaries are squeezed together and forced to have equal slopes at the critical point.
Finally, we note that we did not assume a specific form of $G(T,P,x)$ or $R(T,x)$ in Eqs. \eqref{eq:G_equal} and \eqref{eq:ricci_equal} in our analysis. While our analysis supports the Ruppeiner's $R$-crossing theorem, it also says that it is not only the Ricci curvature of Ruppeiner's metric that can generate a phase boundary consistent with the standard curve. In fact, any construction that chooses pairs of coexistent states lying in the interval $I_g$ and $I_l$ will produce a phase boundary that matches the Maxwell phase boundary at the critical point.

\subsubsection*{Widom line}

In contrast to the Maxwell and Ricci phase boundaries, the agreement of the Widom and Ricci-Widom line near the critical point does not always hold for all our sample systems. In Systems Maria and Rose, the Widom and Ricci-Widom lines run on different directions even at the vicinity of the critical point. System Sina does not have an isobar thermodynamic Widom line, but a Ricci-Widom line can be computed. 

While for the toy systems we observed generally poor agreement of the Widom and Ricci-Widom lines, the complete opposite was seen for the van der Waals case. The Widom and Ricci-Widom lines of the van der Waals fluid are exactly the same, for both the isotherm and the isobar variant. The isobaric heat capacity of the van der Waals fluid can be expressed as\footnote{Normalized variables are used in Eq. \eqref{eq:vdw_CP}. The isobaric heat capacity is nondimensionalized as $C_P\rightarrow (\frac{T_C}{P_C V_C})C_P$, as per Eq. \eqref{eq:cp}.}
\begin{equation}
    C_P(P,x)=\frac{4 \left(5 P+3 x^2(2 x-1) \right)}{3 \left(P+x^2(2 x-3) \right)}.
    \label{eq:vdw_CP}
\end{equation}
Meanwhile, the Ricci curvature of the van der Waals fluid (in the Ruppeiner-$N$ metric) is
\begin{equation}
    R(T,x) = -\frac{x(x-3)^2  \left(x(x-3)^2 -8 T\right)}{8 \left(x(x-3)^2 -4 T\right)^2}.
    \label{eq:vdw_Ricci}
\end{equation}
With the pressure EoS of the van der Waals fluid, the temperature in Eq. \eqref{eq:vdw_Ricci} can be expressed as a function of the pressure and density. While the two functions are completely different, their maxima along isobars and isotherms are exactly coincident. In the work of May and Mausbach\cite{may2012riemannian} where the authors used the usual Ruppeiner-$V$ metric, the Widom and Ricci-Widom lines are only consistent up to their slopes at the critical point, i.e., they are tangent at the point.

It is a curious matter why the Ricci construction of the Widom line works very well for the van der Waals fluid while failing to be consistent with the standard curve for our toy systems. This may be due to the artificial and bare EoS of the toy systems. We may need to impose additional physical constraints in our definition of the free functions in Eq. \eqref{eq:Fform}. For the Lennard-Jones fluid\cite{may2012riemannian}, for example, the Ricci-Widom line was able to approximate the standard Widom line up to having the same slope at the critical point. Knowing when and how the Ricci-Widom line becomes consistent with the standard Widom line is an important question that we intend to investigate in future works.

\section{\label{sec:conclusion}Conclusion and Recommendations}

This work was an exploration of the geodesics and Ricci curvature of the Ruppeiner geometry of different fluid systems. We have discovered that basing geometric prescriptions for important thermodynamic curves, such as phase boundaries and the Widom line, on an alternative Ruppeiner-$N$ metric rather than the usual Ruppeiner metric gives better agreement with standard thermodynamic methods. This alternative Ruppeiner-$N$ metric keeps the number of particles constant, as opposed to the usual Ruppeiner metric which keeps the volume constant. Our overall message is that the Ruppeiner-$N$ metric may be the appropriate metric to use when studying phase diagrams with thermodynamic geometry.

The advantages of the Ruppeiner-$N$ metric already show up in its geodesics. We revisited the pioneering work of Diósi et al.\cite{diosi1989mapping} that first introduced a geodesic-based redefinition of phases for the van der Waals fluid using the standard Ruppeiner-$V$ metric. 
In implementing their classification scheme, however, we found no correspondence between their boundaries and the Widom line in the supercritical region. Moreover, the Diósi boundaries were not even entirely consistent with the standard phase boundary in the subcritical region. This disconnect suggests that their partitioning scheme may have little to do with thermodynamics. In contrast, we have shown that geodesics of the Ruppeiner-$N$ metric detect the presence of the critical isochore, which for the van der Waals fluid is also the isotherm thermodynamic Widom line (i.e., the locus of isobaric heat capacity maxima along isotherms). With our proposed metric, the geodesics select the Widom line as the separating boundary in the supercritical region, thus making thermodynamic geometry more faithful to actual thermodynamics.

Beyond geodesics, we looked at the Ricci curvature of the Ruppeiner-$N$ metric for different fluid systems.
We developed a general expansion of equations of state about a critical point in order to accommodate a broad class of fluid systems in our analysis.
In comparing the phase boundary and the Widom line  generated by the Ricci construction (i.e., $R$-crossing method) to those calculated using standard thermodynamic methods, we found that the phase boundaries are indeed consistent near the critical point. We also provided an explicit proof guaranteeing this agreement for fluids belonging to our EoS family, which unifies the results of May and Mausbach\cite{may2013thermodynamic} and Jaramillo-Gutiérrez et al.\cite{jaramillo2019r}

Finally, we investigated the Ricci construction of the isobar thermodynamic Widom line (i.e., the locus of isobaric heat capacity maxima along isobars). We found that unlike the phase boundary, the standard and Ricci-based Widom lines do not generally agree for systems belonging to our EoS family. This may be due to missing additional physical constraints that should be incorporated in choosing the free functions of our EoS expansion. For more canonical model systems, however, like the van der Waals and the Lennard-Jones fluid\cite{may2012riemannian} the Ricci construction of the Widom line is able to approximate the standard thermodynamic curve close to the critical point. In fact, we proved that for the van der Waals fluid the Ricci construction gives the exact same Widom line as the standard construction, but only when the Ruppeiner-$N$ metric is used. This contrasts with the results of May and Mausbach\cite{may2012riemannian} based on the (usual) Ruppeiner-$V$ metric in which the Ricci-constructed Widom line is only tangent to standard curve at the critical point.

Our work opens several directions for future study. Though we have demonstrated some of the benefits of using the Ruppeiner-$N$ metric, we expect our results to be far from exhaustive. Much of what is known about the Ruppeiner geometry is based on the Ruppeiner-$V$ metric, which is more intuitive and gives a ready interpretation to some quantities that can be computed, but we see no fundamental reason against using the Ruppeiner-$N$ metric. It remains to be seen what other results in the thermodynamic geometry literature will change with our proposed reformulation. In this paper, we have shied away from exploring deeper theoretical and interpretational aspects of the Ruppeiner-$N$ metric, and it remains a mystery why it works better than the traditional Ruppeiner-$V$ metric for the applications we considered.

Finally, we expect further improvements in the EoS expansion we presented in this study. The flexibility of our EoS parametrization is quite wide with five free functions needed to specify a particular thermodynamic system. Physical interpretations are yet to be given to these free functions. It is a curious matter that the Ricci construction should work very well for some systems, but not in general. One potential line of inquiry can start with turning the question around and asking what conditions the EoS must satisfy for the Ricci construction to generate a Widom line that is consistent with the standard thermodynamic one. We leave this and related questions to future work. 

\nocite{*}
\bibliography{aipsamp}

\end{document}